\documentclass[12pt]{article}
\usepackage[mathscr]{eucal}
\usepackage{epsfig,amsfonts}
\usepackage{amsmath}
\usepackage{amsthm,amssymb}
\usepackage{graphicx}
\usepackage{hhline}
\usepackage{cite}
\usepackage{psfrag}
\usepackage{mathrsfs} 
\usepackage{hyperref}
\usepackage{bm}
\usepackage{array}
\usepackage{tikz}
\usepackage{subcaption}
\usepackage{textcomp}
\usepackage{hyperref}
\usepackage{mathrsfs} 
\usepackage{enumitem}
\usepackage{multirow}
\DeclareMathAlphabet{\mathpzc}{OT1}{pzc}{m}{it}

\makeatletter
\@addtoreset{equation}{section}
\makeatother

\def\bea{\begin{eqnarray}}
\def\eea{\end{eqnarray}}
\def\be{\begin{equation}}
\def\ee{\end{equation}}

\def\be{\begin{equation}}
\def\ee{\end{equation}}
\def\bdm{\begin{displaymath}}
\def\edm{\end{displaymath}}
\def\bea{\begin{eqnarray}}
\def\eea{\end{eqnarray}}

\def\ri{{\rm i}}
\def\half{\textstyle\frac{1}{2}}
\def\XXint#1#2#3{{\setbox0=\hbox{$#1{#2#3}{\int}$}
    \vcenter{\hbox{$#2#3$}}\kern-.5\wd0}}

\newcommand{\rd}{\mbox{d}}
\newcommand{\re}{\mbox{e}}

\DeclareMathAlphabet{\mathpzc}{OT1}{pzc}{m}{it}

\usepackage{calc}

\begin{document}
\captionsetup[figure]{labelfont={small},labelformat={default},labelsep=period,name={Fig.\!\!}}
\captionsetup[table]{labelfont={small},labelformat={default},labelsep=period,name={Tab.\!\!}}

\begin{titlepage}
\begin{flushright}
$\phantom{{\it tresrtfdsgqw }}$\\
\end{flushright}
%\begin{flushright}
%RUNHETC-2016- \\
%\end{flushright}

\vspace{0.8cm}

\begin{center}
\begin{LARGE}

{\bf Bethe state norms 
for the Heisenberg spin chain
in the scaling limit
}

\end{LARGE}
\vspace{1.3cm}
\begin{large}

{\bf
% Vladimir V. Bazhanov$^{1}$, 
 Gleb A.  Kotousov$^{1,2}$  
%\bigskip
%Sergii Koval$^{1}$
 and Sergei  L. Lukyanov$^{1,3}$}

\end{large}

\vspace{1.cm}
${}^{1}$NHETC, Department of Physics and Astronomy\\
     Rutgers University\\
     Piscataway, NJ 08855-0849, USA\\
\vspace{.3cm}
$^2$Department of Theoretical Physics\\
         Research School of Physics and Engineering\\
    Australian National University, Canberra, ACT 2601, Australia\\[0.2cm]
%$^2$Mathematical Sciences Institute\\
%      Australian National University, Canberra, ACT 0200,
%      Australia\\\ \\
%\vspace{.1cm}
and\\
\vspace{.2cm}
${}^{3}$Kharkevich Institute for Information Transmission Problems,\\
Moscow, 127994, Russia
\vspace{1.0cm}

\end{center}

\begin{center}
\centerline{\bf Abstract} \vspace{.8cm}

\parbox{13cm}{%
In this paper we discuss the norms  of the Bethe states  for the  spin $\half$  Heisenberg chain in the critical regime.
Our analysis is based on the ODE/IQFT correspondence. 
Together with numerical work, this has lead us to
 formulate  a set of conjectures concerning the scaling behavior of the norms.
Also, we clarify the r${\hat {\rm o}}$le of the different
Hermitian structures associated with the integrable structure studied in the series of works of Bazhanov, Lukyanov and Zamolodchikov in
the mid nineties.
}
\end{center}

\vfill

\end{titlepage}
\setcounter{page}{2}

\section{Introduction}

Consider the  Hamiltonian of the  Heisenberg spin $\half$ chain 
\bea\label{asiisaias}
\hat{{\mathsf H}}=-\sum_{i=1}^L\big(\sigma_i^x\sigma_{i+1}^x+\sigma_i^y\sigma_{i+1}^y+\Delta\, \sigma_i^z\sigma_{i+1}^z\big)\ ,
\eea
subject to the 
 twisted boundary conditions involving the parameter  $-\half<{\tt k}\leq\half $:
\bea\label{kssus}
\sigma^{\pm}_{L+1}=\re^{\pm 2\pi\ri {\tt k}}\ \sigma^{\pm}_{1}\ ,\ \ \ \ \ \sigma_{L+1}^z=\sigma_1^z\ .
\eea
Since it commutes  with the $z$-component of the total spin operator 
%$\half \sum_{i=1}^L{\vec \sigma}_i$,
 they can be simultaneously diagonalized. If we let the $M=\tfrac{1}{2}\,L-S^z$
``down'' spins be at sites
$1\leq x_1<x_2\cdots <x_M\leq L$,
then the (unnormalized)  Bethe Ansatz (BA)  wave function has the form
\bea\label{aisaiasu}
\Psi(x_1,\ldots,x_M)=\sum_{\hat P}A_{\hat P}\ \re^{\ri\sum_{m=1}^M p_{{\hat P}m}x_m}\ .
\eea
Here $\sum_{\hat P}$ stands for the sum over the $M!$ permutations of $1,2,\ldots, M$  and 
$ A_{\hat P}$ is given by the product of ``two-body'' factors:
\bea\label{aisasau}
A_{\hat P}=\prod_{1\leq j< m\leq M}f(p_{{\hat P}j}, p_{{\hat Pm}})\ .
\eea
The quasi-periodicity condition
\bea
\Psi(x_2,\ldots,x_M,x_1+L)=\re^{2\pi\ri{\tt k}}\ \Psi(x_1,x_2,\ldots,x_M)
\nonumber
\eea
 leads to a  set of $M$ equations specifying  the admissible values of $\{p_j\}_{j=1}^M$ known as 
the BA  equations:
\bea\label{BAS}
%\bigg(
%\frac{\sinh(\beta_j+\frac{\ri\gamma}{2})}
%{\sin(\beta_j-\frac{\ri\gamma}{2})}
%\bigg)^L
\re^{\ri p_j L}
%=-\re^{2\pi\ri{\tt k}}\ \prod_{m=1}^M\frac{\sinh(\beta_j-\beta_m+\ri\gamma)}
%{\sinh(\beta_j-\beta_m-\ri\gamma)}
=\re^{2\pi\ri{\tt k}}\ (-1)^{M-1}\ \prod_{m\not= j}\ \re^{-\ri\Theta(p_j,p_m)}
\eea
%\
%&&
%\re^{-2\pi\ri {\tt k}}\ \re^{{\ri p}_jL}\ \prod_{m\not=j}\frac{f(p_m,p_j)}{f(p_j,p_m)}=1\ .
%\eea
%\bea
%\bigg(
%\frac{\sinh(\beta_j+\frac{\ri\gamma}{2})}
%{\sin(\beta_j-\frac{\ri\gamma}{2})}
%\bigg)^L
%\re^{\ri p(\beta_j) L}
%=-\re^{2\pi\ri{\tt k}}\ \prod_{m=1}^M\frac{\sinh(\beta_j-\beta_m+\ri\gamma)}
%{\sinh(\beta_j-\beta_m-\ri\gamma)}
%=\re^{2\pi\ri{\tt k}}\ (-1)^{M-1}\ \prod_{m\not= j}\ \re^{-\ri\Theta(\beta_j-\beta_m)}\ ,
%\eea
%where
%\bea\label{aisaisaias}
%\re^{-\ri \Theta(\beta)}=\frac{\sinh(\ri{ \gamma}+\beta)}{\sinh(\ri{ \gamma}-\beta)}\ .
%=
%\frac{1}{\ri}\ \log\bigg(\frac{\sinh(2\ri{ \eta}+\beta)}{\sinh(2\ri{ \eta}-\beta)}\bigg)
%\eea
with
\bea\label{aosaosai}
%\varepsilon(p)=p L-2\pi{\tt k}+\sum_{m=1}^M\log\Theta(p, p_m)\ ,\ \ \ \ \ 
\re^{\ri \Theta(p_j,p_m)}=-\frac{f(p_m, p_j)}{f(p_j,p_m)}\ .
\eea

The subject of our  interest   is the normalization sum
\bea
{\mathfrak  N}=\sum_{1\leq  x_1<x_2<\ldots<x_M\leq L}
\big|\Psi(x_1,\ldots,x_M)\big|^2\ .
\eea
%which can be considered as a functional defined  on  the set of Bethe states.
In ref.\!\cite{Gaudin:1981cyg} strong arguments were presented  to justify the   formulae
%\bea\label{aiusauasss}
%{\mathfrak   N}=\prod_{j=1}^M\frac{1}{\big(-\partial_\beta p(\beta_j)\big)}\ 
%\det\bigg(
%\frac{\partial^2 Y}{\partial\beta_j\partial \beta_l}\bigg)\ 
%\prod_{j\not= m}^M f(p_{m},p_j)\ 
%\ .
%\eea
\bea\label{aiusauasssu}
{\mathfrak   N}=
\det ({\boldsymbol  Y})\ 
\prod_{1\leq j< m\leq M} | f(p_{m},p_j)|^2\ 
\ ,
\eea
where the elements of the $M\times M$ matrix ${\boldsymbol  Y}$ are expressed in terms of $\Theta(p_j,p_m)$ \eqref{aosaosai}
\bea
({\boldsymbol Y})_{jm}=\bigg(L+\sum_{n=1}^M\partial_{p_j}\Theta(p_j,p_n)\bigg)
\ \delta_{jm}-\partial_{p_j}\Theta(p_j,p_m)\ .
\nonumber
\eea
The formula \eqref{aiusauasssu}  was inspired by the original Gaudin hypothesis \cite{Gaudin:1972} for the Lieb-Liniger model.
Its  rigorous proof was given by Korepin  %in the framework of the Quantum Inverse Scattering Method 
in ref.\!\cite{Korepin:1982ej}.

\bigskip
The Bethe wave function contains an ambiguity related to the choice of the two-body  factor $f(p_j,p_m)$. Indeed, 
 the substitution
\bea
f({p_j,p_m})\mapsto g(p_j)g(p_m)\ f({p_j,p_m})\ ,\nonumber
\eea
where $g(p)$  is an arbitrary function of $p$, does not affect the BA equations \eqref{BAS}. However this  changes the 
overall normalization of the wave function \eqref{aisaiasu}\,\eqref{aisasau}
%\bea
%\Psi_L(x_1,\ldots,x_M)\mapsto \Big(\prod_{j=1}^m g(p_j)\Big)^{M-1}\ \Psi_L(x_1,\ldots,x_M)\ ,
%\eea 
and, in  turn, affects    the norm ${\mathfrak N}$.
In what follows we will assume a certain  choice   for $f(p_j,p_m)$. 
To describe it  explicitly 
let us parameterize  the anisotropy  
as
\bea
\Delta=\tfrac{1}{2}\,(q+q^{-1})\nonumber
%-\cos(\gamma)
%=\cos(2\eta)
\eea
and substitute
$\{\re^{\ri p_j}\}_{j=1}^M$   by the set
$\{\zeta_j\}_{j=1}^M$:
\bea\label{asisaisa}
p_j=-\ri\, \log\bigg(\!\frac{1+q\,\zeta_j}{q+\zeta_j}\!\bigg)
%\re^{\ri p(\beta_m)}\ , \  \ \ \ \ \ \ \ \ \ \ \ \ \ \ \ \  \ \ p(\beta)=\frac{1}{\ri}\ 
%\log\bigg(\frac{\sinh(\frac{\ri\gamma}{2}+\beta)}
%{\sinh(\frac{\ri\gamma}{2}-\beta)}\bigg)
%=
%\frac{\cosh(\beta-\ri\eta)}
%{\cosh(\beta+\ri\eta) }
\ .
\eea
With this parametrization
\bea\label{aisaisaias}
\Theta(p_j,p_m)=-\ri\,  \log\bigg(\!\frac{q\,\zeta_m-q^{-1}\zeta_j}{q\,\zeta_j-q^{-1}\zeta_m}\!\bigg)
%\theta(p_j,p_m)=\Theta(\beta_j-\beta_m)\ ,\ \ \ \ \ \ \
%\Theta(\beta)=\frac{1}{\ri}\ \log\bigg(\frac{\sinh(\ri{ \gamma}-\beta)}{\sinh(\ri{ \gamma}+\beta)}\bigg)\ ,
%=
%\frac{1}{\ri}\ \log\bigg(\frac{\sinh(2\ri{ \eta}+\beta)}{\sinh(2\ri{ \eta}-\beta)}\bigg)
\eea
so that
the  two-body factors in   \eqref{aisasau} can be chosen  as
\bea\label{kaajay}
f(p_{j},p_{m})=\frac{q\,\zeta_j-q^{-1}\zeta_m}{\zeta_m-\zeta_j}\ .
%\frac{\sinh(\beta_j-\beta_m+\ri\gamma)}{\sinh(\beta_j-\beta_m)}\ .
%=-\frac{\sinh(\beta_j-\beta_m-2\ri\eta)}{\sinh(\beta_j-\beta_m)}
%\\
%\frac{A_{m,j}}{A_{j,m}}&=&\re^{\ri\Theta(\beta_j-\beta_m)}\ \ \ \ \ \ {\rm with}\ \ \ \ \  \ \
 %\Theta(\beta)=\frac{1}{\ri}\ \log\bigg[\frac{\sinh(\ri\gamma-\beta)}{\sinh(\ri\gamma+\beta)}\bigg]
\eea

\bigskip
%We consider the case $0\leq\gamma<\pi$,
%when the spin chain is critical.
In the last thirty years     impressive  progress   has been made
 for the   exact calculation of a variety of interesting
quantities in the Heisenberg model. However,
 to the best of our knowledge, we are still lacking
a simple analytical tool to study the large-$L$ behavior of the norms $\mathfrak{N}$
 \eqref{aiusauasssu}-\eqref{kaajay}.
We consider the case with $|q|=1$, when the spin chain is critical.
Then
our numerical analysis suggests that 
the norms
  corresponding to the low energy excitations  possess  the following large-$L$   behavior
\bea\label{aososaik}
{\mathfrak  N}={\mathfrak N}_\infty\ L^{\eta}\ \re^{{\cal A}_2 L^2+{\cal A}_1 L}\ \big(1+o(1)\big)\ .
\eea
It is not difficult to find the exact expressions for the coefficients ${\cal A}_{1,2}$ 
that define the leading large-$L$ asymptotic.
Moreover, as the numerical calculation of ${\mathfrak N}$  
for the chain with  $100\div 1000$ sites requires a negligible amount of
  computer time nowadays, it is possible to easily  determine the  scaling exponent $\eta$ and amplitude $ {\mathfrak N}_\infty$
   with a relative accuracy of 
$ 10^{-4}\div 10^{-5}$
as long as the anisotropy parameter  $\Delta$ is not too close to $\pm1$.
With the numerical data at hand,
one can try to guess the analytical expression for $\eta$ and ${\mathfrak N}_\infty$ in  \eqref{aososaik}.
The main purpose of this work is to formulate a set of conjectures concerning the form of these
scaling quantities.

 \newpage
 
 \section{RG flow of the low energy   Bethe states}

Some immediate clarifications
are needed for eq.\,\eqref{aososaik}. 
In assigning an $L$ dependence to the norms, we have in mind a family of Bethe states
$|\Psi_L\big\rangle$ defined
for different lengths of the spin chain.
For a general  lattice system, there are difficulties in introducing  the $L$-dependence (RG flow)
for individual stationary states. Of course, 
 since the Hilbert space is not isomorphic for different lattice sizes,
the problem  only makes sense for the low energy part of the spectrum. 
It is
clear how to assign such a  dependence for the ground state  or, for that matter, the lowest energy
states in the disjoint sectors of the Hilbert space. 
However 
 forming  individual   RG flow trajectories 
 for low energy stationary states that are 
densely distributed does not seem to be a trivial task.

\bigskip

In the case under consideration the problem is greatly facilitated by
the existence of the BA equations,  which are useful to re-write in the logarithmic form
\bea\label{klkwqnmsd}
L p_j=2\pi{\tt k}-2\pi\, I_j-\sum_{m=1}^M \Theta(p_j, p_m)\ .
\eea
Here  
$I_j$ are the so-called Bethe numbers which are integers or half-integers for $M$ odd or even respectively. An unambiguous definition of $I_j$  requires fixing the
branches of the logarithms in the formulae \eqref{asisaisa},\,\eqref{aisaisaias}.
Although this  is an important step in  any practical calculation, we will not touch on it here
and only mention  that 
\bea\label{kssusu}
I^{\rm (vac)}_j=-\tfrac{1}{2}\ (M+1)+j\ \ \ \ \ \ \ \ \ \ \ \big(\, j=1,\ldots, M=\tfrac{1}{2}\,L-S^z\,\big)
\eea
for the vacuum state in the sector with fixed value of $S^z$.
For sufficiently large $L$ the Bethe numbers corresponding to the low  energy states are given
by $I^{\rm (vac)}_j+\delta I_j$, where the variation $\delta I_j$ from the ``vacuum'' distribution \eqref{kssusu}
are nonzero only in the vicinity  of the  edges, i.e., for $j\ll M$  or $M-j\ll M$. 
The set  $\{\delta I_j\}$ can  be used to define
the individual  RG flow trajectories  $|\Psi_L\rangle$ with the following procedure.

Starting with a spin chain for relatively small $L$ one can perform the
numerical diagonalization of the Hamiltonian.
The latter is part of a family of commuting operators,
of which a prominent r$\hat{{\rm o}}$le is played by the so-called $Q$-operator:
%(see Appendix for
%details):
\bea
[{\hat {\boldsymbol {\mathsf Q}}}(\zeta), {\hat {\mathsf H}}]=0\ .
\nonumber
\eea
Together with the energies, one should compute the corresponding 
eigenvalues of $\hat{\boldsymbol{{\mathsf Q}}}$ which are polynomials,
 \begin{equation}\label{Qeig1}
{\mathsf Q}(\zeta)\,=\, \prod\limits_{j=1}^{M}\,({\zeta}_j-{\zeta})\ ,
\end{equation}
 whose zeros
$\{\zeta_j\}_{j=1}^M$ are related  to  $\{p_j\}_{j=1}^M$ through   eq.\eqref{asisaisa}.
This allows one to extract the Bethe roots for  a certain Bethe state $|\Psi_L\rangle$
and, using \eqref{klkwqnmsd}, also the
set $\{\delta I_j\}$.  For the state $|\Psi_{L+2}\rangle$,
the BA equations are specified to have the same  $\{\delta I_j\}$.
Moreover, for their iterative solution the initial approximation
can be constructed using the Bethe roots for $|\Psi_L\rangle$.
This procedure provides a way for defining
the RG flow of an individual Bethe state. Finally, the norm
entering in the l.h.s. of  eq.\,\eqref{aososaik} is understood as
\bea\label{aosaiosia}
\mathfrak{N}=\langle\Psi_L\,|\,\Psi_L\rangle\ .
\eea

 \section{Scaling behavior of the Bethe roots}

 The BA equations \eqref{BAS}, being written for the set $\{\zeta_j\}_{j=1}^M$ \eqref{asisaisa}, form   an algebraic system\footnote{In this paper it is  always 
 assumed  that $S^z\geq 0$, i.e., $M\leq \half\,L$.}
\begin{equation}\label{AB1}
\bigg(\!\frac{1+q\,  \zeta_j}{1+q^{-1}\zeta_j}\!\bigg)^L\,=\,-\re^{2\pi\ri {\tt k}}\ q^{2S^z}\,
\prod\limits_{m=1}^{\frac{L}{2}-S^z}\,\frac{\zeta_m-q^2\, \zeta_j}{\zeta_m-q^{-2}\zeta_j}\ \ \ \ \ \ \ \ \ \ \ 
\big(j=1,\ldots, \half L-S^z\big)\ .
\end{equation}
 In the case of the  $XXZ$ spin chain  in the critical regime,  any set 
solving this algebraic system coincides with its complex conjugate:\footnote{This follows
from the conjugation condition for the $Q$-operator $\hat{\boldsymbol{\mathsf{Q}}}^\dagger(\zeta)=
\hat{\boldsymbol{\mathsf{Q}}}(\zeta^*)$, which implies that
all  the eigenvalues ${ \mathsf{Q}}(\zeta)$ \eqref{Qeig1} are real analytic functions of $\zeta$.}
%Despite
%that this property is
%commonly known and used, we lack a rigorous proof.}
\be\label{asdi1231}
\{\zeta_j\}_{j=1}^M=\{\zeta^*_j\}_{j=1}^M\ .
%\nonumber
\ee
This important property allows one 
to rewrite eq.\eqref{aiusauasssu},\,\eqref{kaajay} in the form
\bea\label{aiusssu}
{\mathfrak   N}=
\det ({\boldsymbol Y})\ 
\prod_{j\not=m}   \frac{q\,\zeta_j-q^{-1}\zeta_m}{\zeta_m-\zeta_j}\ 
\ .
\eea

\begin{figure}
%[h]
%\includegraphics[width=\textwidth]{vacroots}
\centering
\vskip-2cm
\scalebox{1.}{
\includegraphics{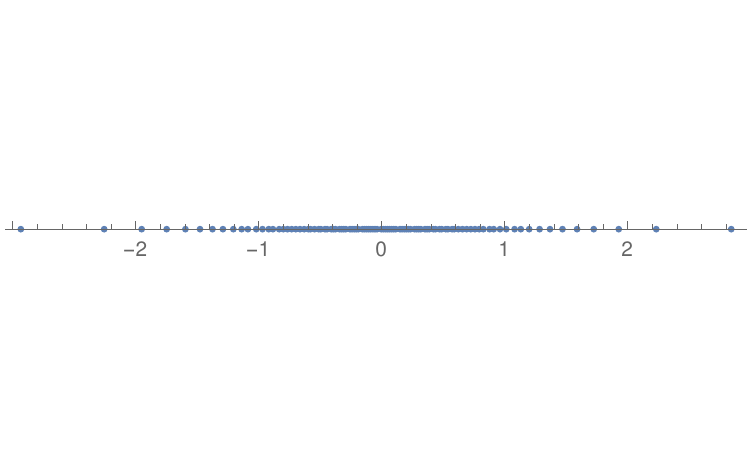}
}
\vskip-3cm
\caption{\small \!The distribution of $\theta_j=-\frac{1}{2}\ \log (\zeta_j)$ 
for the ground state in the sector $S^z\,=\,0$, for $L=200$, $\Delta=\cos(\frac{2\pi}{5})$ and $\pi{\tt k}\,=\,\frac{1}{10}$.\label{figb2111r}}
\end{figure}

Let us now give a short sketch of how one can obtain the leading large $L$ asymptotic behavior ${\mathfrak N}\sim \re^{{\cal A}_2L^2}$.
 First consider the vacuum
 in the sector with given spin $S^z$. %\footnote{In this work we always  assume that the chain has an even number of sites.}
For such a state, if the twist parameter ${\tt k}$ 
is sufficiently small,   all the Bethe roots $\zeta_j$ are real and positive.
Using the BA equations one can show that the numbers
\bea
\theta_j\equiv -\half\ \log(\zeta_j)
\nonumber
\eea
are distributed 
within the segment $[\Lambda_-,\Lambda_+]$ with $|\Lambda_\pm|\propto \log(L)$ (see fig.\ref{figb2111r}). As $L\gg 1$  
 most of the roots become densely packed
about the origin
so that 
\bea\label{uytu678}
\rho_L(\theta_{j+\frac{1}{2}})=\frac{2}{L(\theta_{j+1}-\theta_j)}\ \ \ \ \ \ \ \ {\rm with}\ \ \ \theta_{j+\frac{1}{2}}=\tfrac{1}{2}\ \big(\theta_{j+1}+\theta_j\big)
\eea
\begin{figure}
\centering
\scalebox{0.8}{
\begin{tikzpicture}
\node at (0,3.6) {$\rho(\theta)$};
\node at (5.4,-2.8) {$\theta$};
\node at (0,0) {\includegraphics[width=10cm]{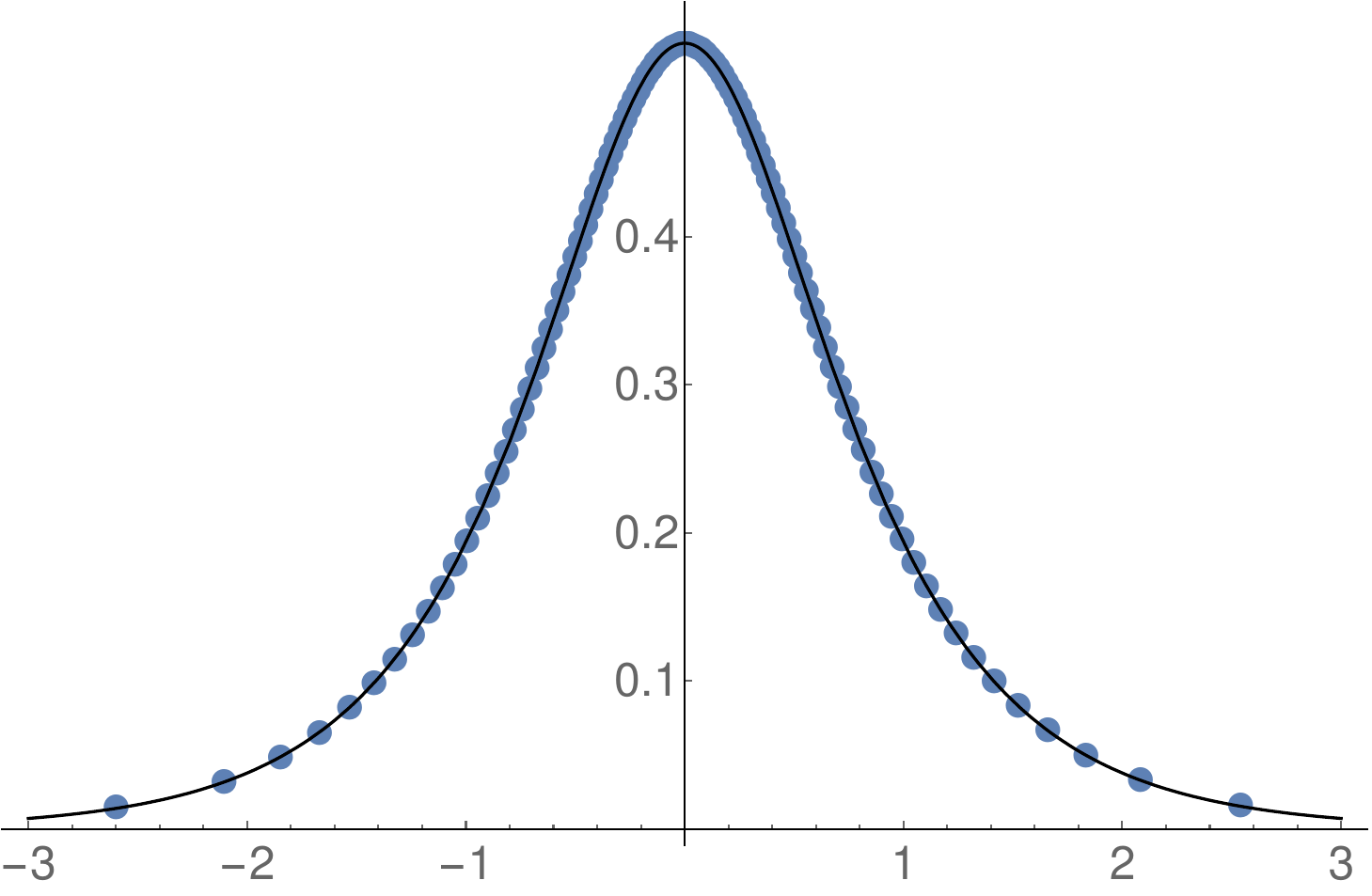}};
\end{tikzpicture}
}
\caption{\small \!A plot of the continuous density $\rho(\theta)=\frac{1}{\pi(1-\beta^2)\cosh(\frac{\theta}{1-\beta^2})}$
for $\beta^2=\frac{2}{5}$ and $\pi{\tt k}=\frac{1}{10}$. The dots represent $\rho_L$ \eqref{uytu678} obtained from
the Bethe roots shown in fig.\,\ref{figb2111r}.
\label{figa2}}
\end{figure}
becomes the continuous density  and (see  fig.\ref{figa2}) 
\bea\label{iasisai}
\lim_{L\to\infty}\rho_L=\frac{1}{\pi(1-\beta^2)\cosh(\frac{\theta}{1-\beta^2})}\ .
%\nonumber
\eea
Here we use the parameter
\bea
\beta\ :\ \ \ q=\re^{\ri\pi \beta^2}\ ,\ \ \ \   0< \beta\leq 1\ .
\nonumber
\eea
With these well known facts (see e.g.\cite{Baxter:1982zz})
it is easy to show that the leading asymptotic behavior of the 
logarithm of the double  product  in   the r.h.s. of   \eqref{aiusssu} grows as
${\cal A}_2\,L^2$ with 
\bea\label{aisiasjjs}
{\cal A}_2=\int_{0}^\infty\frac{\rd t}{t}
\ \frac{\sinh( \frac{\beta^2 t}{1-\beta^2} )\ \sinh(t)}{2\sinh(\frac{t}{1-\beta^2})\, \cosh^2(t)}\ .
\eea
At the same time
one can argue that $\det
({{\boldsymbol Y}})$ in \eqref{aiusssu}  goes as $L^L$, i.e, 
 does not contribute to the leading large $L$ asymptotic of the norm.
 Finally one notes that for  the low energy excitations with $L\gg 1$,
   the pattern of the Bethe roots are changed essentially only in the 
 vicinity of the edges and the density of the majority of the roots  is still given by \eqref{iasisai}.
As a result, the leading large-$L$  asymptotic of their norms remains the same.
%In a similar way, one can find the value of the constant  ${\cal A}_-$ given by the second line in \eqref{aisias}.

 %Contrary to the leading large $L$ behavior of ${\mathfrak N}$,
 
 It is a more difficult task to justify the full asymptotic formula
 \eqref{aososaik}. Nevertheless, numerical work supports this relation. It also 
shows that the constant ${\cal A}_1$   remains the
 same for the low energy states and reads explicitly as
 \bea\label{asdasd1122}
 {\cal A}_1=\log\bigg(\!\frac{2(1-\beta^2)}{\sin(\pi \beta^2)}\!\bigg)\ .
 \eea
 At the same time both the exponent  $\eta$ and the amplitude ${\mathfrak N}_\infty$ do depend on the state $|\Psi_L\rangle$, i.e,
 their value is influenced by the  roots near  the edges of the locus of  the Bethe roots distribution.

 \bigskip
 
 In the case of the vacuum states, where the Bethe numbers are given by \eqref{kssusu}, all the roots $\zeta_j$ are real and ordered as
 $\zeta_1<\zeta_2<\ldots<\zeta_M$. For the excited states the roots may become complex and can be ordered w.r.t. their real part
 \bea
 \Re e(\zeta_1)\leq \Re e(\zeta_2)\leq\ldots  \leq \Re e(\zeta_M)\nonumber
 \eea
 (the ordering prescription for  the Bethe roots with coinciding real parts is not essential for us at this point).
When both $L$ and $M=\tfrac{1}{2}\,L-S^z$ go to infinity  with $S^z$ kept fixed, 
the roots  $\zeta_j\to  0$ and $\zeta_{M-j}\to  \infty$.
More accurately for given $j=1,2,\ldots$ there exists the following limits
 \bea\label{jassasay}
 s_j=\lim\limits_{L\to\infty}\big(\tfrac{L}{\pi}\big)^{2(1-\beta^2)}\,\zeta_j\ ,\ \ \ \ \ \ \ {\bar s}_j=\lim\limits_{L\to\infty}
 \big(\tfrac{L}{\pi}\big)^{2(1-\beta^2)}\, (\zeta_{M-j})^{-1}\ .
 \eea
 In these formulae we have used $L/\pi$ instead of $L$ due to the following reason. For $1\ll  j\ll L$ one can approximate the
 Bethe roots by $\zeta_j\approx (L/\pi)^{-2(1-\beta^2)}\,s_j$  and $\zeta_{M-j}\approx 
 (L/\pi)^{+2(1-\beta^2)}\,
 ({\bar s}_{j})^{-1}$.
It is easy to check that these formulae are consistent with the relations \eqref{uytu678},\eqref{iasisai}, provided that
\be\label{jssus}
(s_j)^{\frac{1}{2(1-\beta^2)}}=j+O(1)\,,\qquad (\bar{s}_j)^{\frac{1}{2(1-\beta^2)}}=j+O(1)\ \ \ \ \ (j\to\infty)\ .
\ee 
 This simple form of the asymptotics  justifies the appearance of $\pi$ 
entering into the definition of the scaling limit \eqref{jassasay} of the Bethe roots.
 
 \bigskip
 
 \section{BA equations in the scaling limit}

Keeping in mind the scaling behavior \eqref{jassasay}, one can consider
 the scaling limit of  the BA equations \eqref{AB1}.
 To simplify the discussion, we make the technical assumption that $0<\beta^2<\frac{1}{2} $. 
 In this case, as it follows from \eqref{jassasay}, 
 $\zeta_j$  with $j$ fixed decays faster  than $L^{-1}$ so that the l.h.s. of \eqref{AB1} turns to be one  as $L\to\infty$.
 Hence, the BA equations take the form
 \bea\label{aisiasiss}
\re^{2\pi\ri {\tt K}}\ \prod\limits_{n=1}^{\infty}\,\frac{1-q^{+2} s_j/s_n}{1-q^{-2}s_j/s_n}=-1\ ,
\eea
where
\bea\label{saisisas}
\re^{2\pi\ri {\tt K}}=q^{S^z}\ \re^{+2\pi\ri{\tt k}}\ .
\eea
 Similarly, 
we can consider \eqref{AB1}  assuming that 
${\bar j}\equiv M-j$ is
 kept fixed
as $L\to\infty$. 
This leads to equations
  for the set $\{{\bar s}_j\}_{j=1}^\infty$,
which differ from \eqref{aisiasiss} only in nomenclature:
 \bea\label{asiiasisa}
 \re^{2\pi\ri {\bar {\tt K}}}\ \prod\limits_{n=1}^{\infty}\,\frac{1-q^{+2} \,{\bar s}_j/{\bar s}_n}
{1-q^{-2}\,{\bar s}_j/{\bar s}_n}=-1
\eea
with
\bea\label{aiaisasa}
\re^{2\pi\ri{\bar {\tt K}}}=
q^{S^z}\ \re^{-2\pi\ri {\tt k}}\ .
\eea
 Finally eqs.\eqref{aisiasiss} and \eqref{asiiasisa} can be rewritten in logarithmic form
 \bea\label{jaasusa}
 \frac{1}{2\pi\ri}\ \log\bigg[\frac{A(q^{+2} s_j)}{A(q^{-2}s_j)}\bigg]=\frac{1}{2}-{\tt K}-{m}_j\ ,\ \ \ \ 
  \frac{1}{2\pi\ri}\ \log\bigg[\frac{{\bar A}(q^{+2} {\bar s}_j)}{{\bar A}(q^{-2}{\bar s}_j)}\bigg]=\frac{1}{2}-
  {\bar {\tt K}}-{\bar { m}}_j\ .
 \eea
 Here we use the notations
 \bea\label{issus1a}
A(s)=\prod_{n=1}^\infty\Big(1-\frac{s}{s_n}\Big)\ ,\ \ \ \ \ \ {\bar A}(s)=\prod^\infty_{n=1}\Big(1-\frac{s}{{\bar s}_n}\Big)\ ,
\eea
where the convergence of the products are guaranteed by the conditions \eqref {jssus} provided that 
$0<\beta^2<\half$.
 The low energy Bethe states are distinguished by a specific phase agreement in \eqref{jaasusa}, determined
 by a choice of integers $\{{ m}_j\}_{j=1}^\infty$ and $\{{\bar { m}}_j\}_{j=1}^\infty $ which play
 the r${\hat {\rm  o}}$le  similar to that of the Bethe numbers $I_j$ in \eqref{klkwqnmsd}.
 These integers of course depend on the 
 choice of branches of the logarithm in \eqref{jaasusa}. However, once these branches are suitably fixed, every 
 low energy state is characterized by  the two  sets $\{{m}_j\}_{j=1}^\infty$ and $\{{\bar { m}}_j\}_{j=1}^\infty $.
 
\begin{figure}
\centering 
\begin{tikzpicture}
\node at (-4.5,0) {\includegraphics[width = 0.46\textwidth]{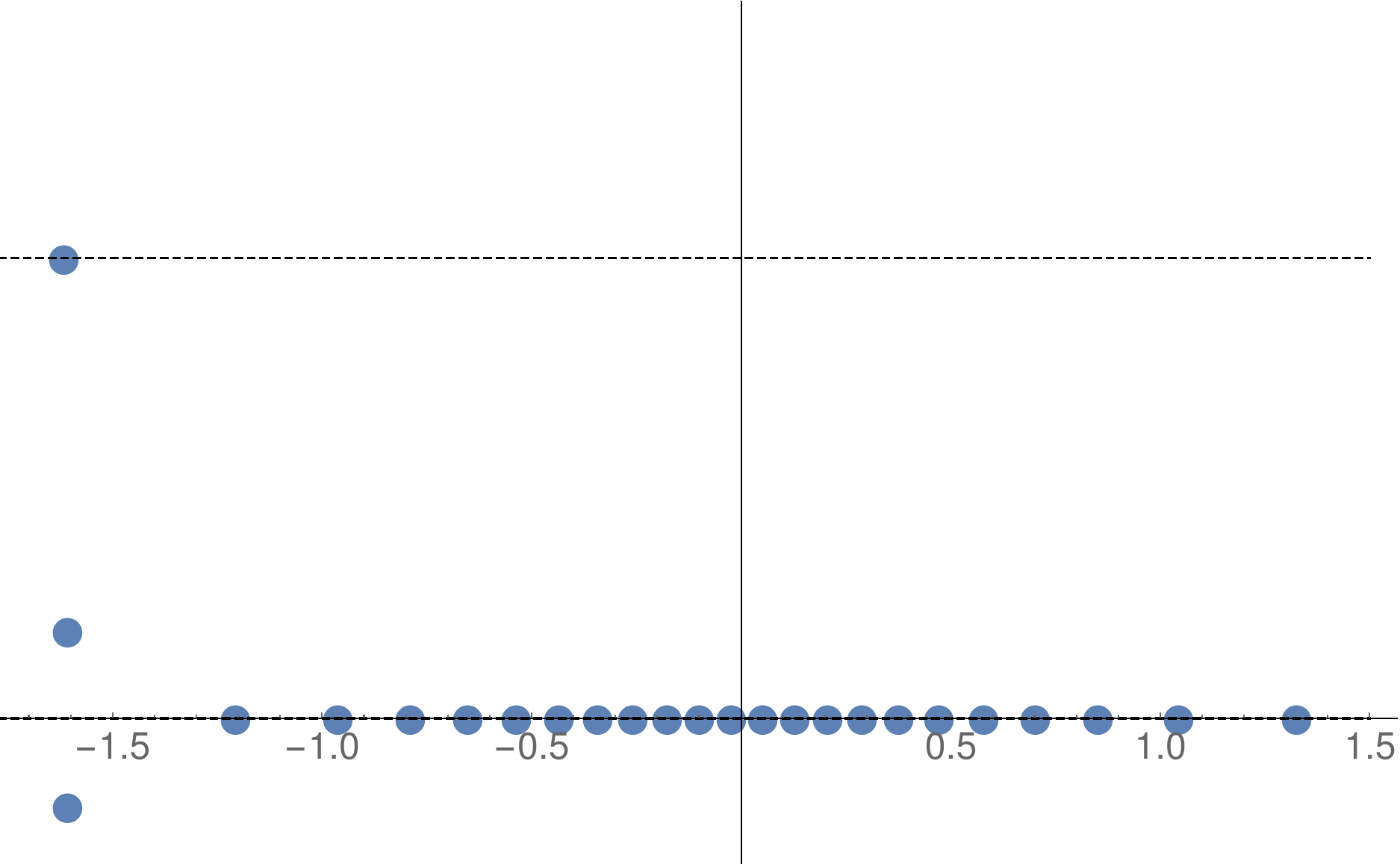}};
\node at (4.5,0) {\includegraphics[width = 0.46\textwidth]{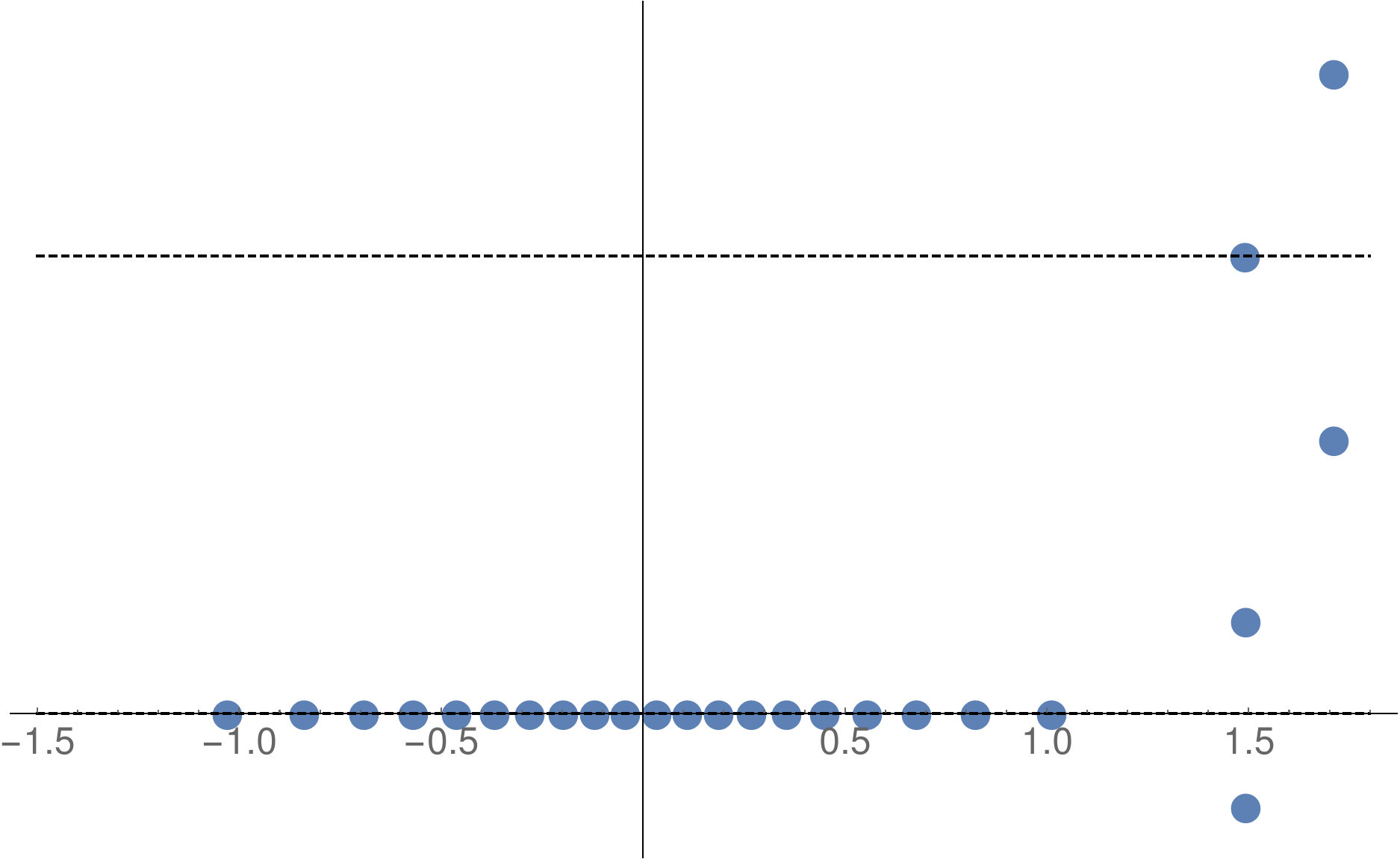}};
\node at (-6.5,1.25) {\small $\Im m(\theta)=\frac{\pi}{2}$};
\node at (2,1.25) {\small $\Im m(\theta)=\frac{\pi}{2}$};
\node at (-3,2.05) {$\theta$};
\node at (5.5,2.05) {$\theta$};
\draw (-3,2.05) circle (0.3cm);
\draw (5.5,2.05) circle (0.3cm);
\end{tikzpicture}
\caption{\small \!The distribution of $\theta_j$ in the complex plane $\theta=-\frac{1}{2}\ \log (\zeta)$
for the primary winding states  with  the winding numbers ${\tt w}=+1$
(left panel)  and ${\tt w}=-2$ (right panel). The value of the remaining parameters is
$S^z\,=\,0$, $L=50$, $\Delta=\cos(\frac{2\pi }{5})$ and $\pi{\tt k}\,=\,\frac{1}{10}$.
  \label{figa3}}
\end{figure}
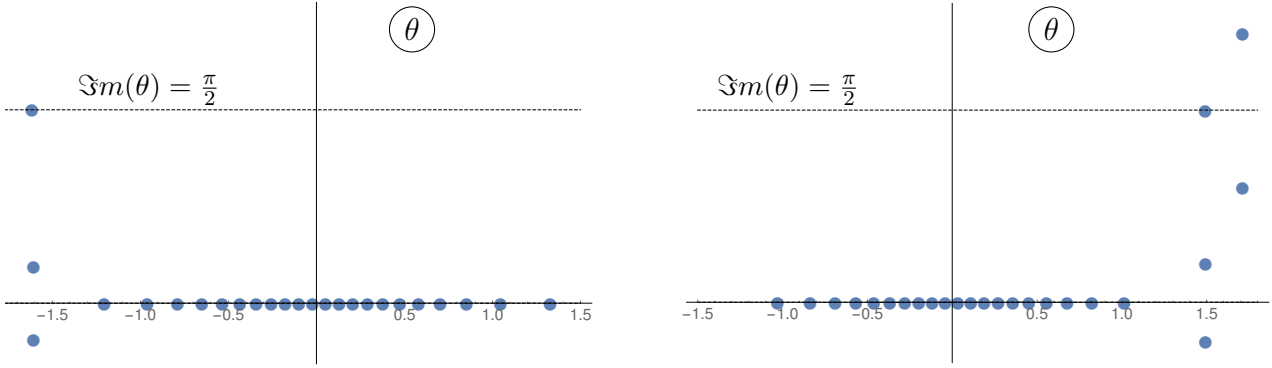

 \bigskip
 The parameters ${\tt K}$ and ${\bar {\tt K}}$ which show up in \eqref{jaasusa} call for some explanation.
 Since the Hamiltonian of the $XXZ$ spin chain \eqref{asiisaias},\,\eqref {kssus} is a periodic function of the twist parameter  ${\tt k}$,
 relations \eqref{saisisas} and \eqref{aiaisasa} imply that
 \bea
 {\tt K}=\beta^2\,S^z
 +({\tt k}+{\tt w})\ ,\ \ \ \ \ \ {\bar {\tt K}}=\beta^2\,S^z-({\tt k}+{\tt w})\,,
 \eea
 where ${\tt w}=0,\pm 1,\pm 2,\ldots$\ .
Below  we will refer to the integer ${\tt w}$ as a winding number. It enumerates the
different bands of the spectrum of the model.
The lowest energy states in  the sector with given ${\tt w}$ 
are of special interest.
We will call these the primary Bethe states.
The patterns of the Bethe roots  are depicted
for two such states corresponding to the winding numbers ${\tt w}=+1$
and ${\tt w}=-2$ in fig.\,\ref{figa3}.

\bigskip

 Finally let us note that the functions $A(s)$ and  ${\bar A}(s)$ \eqref{issus1a} can be interpreted as a properly defined
 scaling limit of the eigenvalues of the $Q$-operator. 
Indeed, as it follows from eqs.\eqref{Qeig1},\,\eqref{jassasay}
 \bea\label{oasidoi192}
 A(s)=\lim_{L\to\infty} \frac{ {\mathsf Q}(\zeta )}{{\mathsf Q}(0)}\bigg|_{\zeta=(L/\pi)^{-2(1-\beta^2)}\,s}\ \, ,\ \ \ \ \ 
 {\bar A}(s)=\lim_{L\to\infty} (-\zeta)^{-M} {\mathsf Q}(\zeta )\big|_{\zeta=(L/\pi)^{+2(1-\beta^2)}\,s^{-1}}\ .
 \eea
Notice that in writing eq.\eqref{Qeig1}   the  $Q$-operator was normalized by the condition
\bea
 \lim_{\zeta\to\infty} (-\zeta)^{{\hat {\mathsf S}}^z-\frac{L}{2}}\ {\hat{\boldsymbol{\mathsf Q}}}(\zeta)={\hat 1}\ .
 \eea
In this case
 \bea
{\mathsf Q}(0)=\prod_{j=1}^M\zeta_j\ .
 \eea

\section{ODE/IQFT correspondence}

It is well known that the scaling behavior of the Heisenberg spin chain
is governed by the massless Gaussian model with the Lagrangian density
\bea\label{ososos}
{\cal L}=\frac{1}{4\pi}\ (\partial_a \phi)^2\ .
%\nonumber
\eea
Below we use the complex Euclidean coordinates $z=x_2+\ri\, x_1$, ${\bar z}=x_2-\ri\, x_1$
and denote $\partial=\partial/\partial z$,  ${\bar \partial}=\partial/\partial {\bar z}$.
We'll take $x_1\in [\,0,R\,]$
and due to the scale invariance, one can set $R=2\pi$ without any loss of generality.
The equations of motion imply that $\partial\phi$ and 
${\bar \partial}\phi$
are holomorphic and antiholomorphic   fields, respectively, admitting
the Fourier series  expansions
\bea\label{issss}
\partial\phi=-\ri\, \sum^\infty_{m=-\infty} a_m\ \re^{-m z}\ ,\ \ \ \  \ 
{\bar \partial}\phi=-\ri \sum^\infty_{m=-\infty} {\bar a}_m\ \re^{-m {\bar z}}\ .
\eea
The expansion coefficients obey  the
Heisenberg commutation relations,
\bea\label{sisisis}
[a_m,a_n]=[{\bar a}_m,{\bar a}_n]=\tfrac{m}{2}\,\delta_{m+n,0}\ ,\ \ \ \ \ [a_m,{\bar a}_n]=0\ .
\eea
The low energy states of the spin chain
in the sector with given $S^z$
form the linear space
\bea\label{ososos232}
%{\cal H}_{S^z}=
%\sum\limits_{{\tt w}=-\infty}^\infty\!\!\! \!\!\!\!\!\! \!\!\oplus\,
\bigoplus_{{\tt w}=-\infty}^{+\infty}\, {\bar {\cal F}}_{\bar{ p}}\otimes{ {\cal F}}_{ p}
\big|_{{ 2p}=\beta\,S^z
 +({\tt k}+{\tt w})/\beta\atop 2{\bar p}=\beta\,S^z-({\tt k}+{\tt w})/\beta}
 %\nonumber
\eea
in the scaling limit.
Here ${\cal F}_{p}$ and $\bar{{\cal F}}_{\bar{ p}}$ stand for the Fock spaces.
The Fock space  ${\cal F}_{ p}$ is
the highest weight representation
with the highest vector $|{p}\rangle$, defined by the conditions
\bea
a_{n}\,|{ p}\rangle=0\ \ \ \ (\forall \ n>0)\ , \ \ \ \ \ \ \ \ a_0\,|{p}\rangle={p}\,|{ p}\rangle\ .
\nonumber
\eea
Similarly, the Fock space $\bar{{\cal F}}_{\bar{p}}$ is the space of the representation of
another copy of the Heisenberg algebra generated by $\{\bar{a}_n\}$.
\bigskip

On the other hand, our analysis in the previous section
suggests that the low energy states
are characterized by the value of $S^z$, the winding number ${\tt w}$ and
two sets of integers ${ m}_j$ and ${\bar {m}}_j$.
%which,
%for the primary Bethe states are given by eq.\,\eqref{oaisoi1}.
As it follows from eq.\eqref{kssusu},
for the vacuum state with ${\tt w}=0$,  the integers ${ m}_j$ and ${\bar {m}}_j$
are consecutive positive numbers
\bea\label{oaisoi1}
{ m}_j^{(\rm vac)}={\bar {m}}_j^{(\rm vac)}=j\ \ \ \  \ \ \ \ \ \ \ \ (j=1,2,\ldots)\ .
\nonumber
\eea
For a general low energy state the ${m}_j$, ${\bar { m}}_j$
differ from their ``vacuum'' values, but
 stabilize to these for sufficiently large $j$:
\bea
{ m}_j={\bar { m}}_j=j\ \ \ \  \ \ \ \ \ \ {\rm for}\ \ \ j\gg 1\ .
\nonumber
\eea
In this section we will briefly explain the link between the two descriptions.
Our discussion is based on the ODE/IQFT  (Ordinary Differential Equations/Integrable Quantum Field Theory)
correspondence, which was developed
in the works \cite{Dorey:1998pt,Bazhanov:1998wj,Suzuki:2000fc,Bazhanov:2003ni}.
\bigskip

In ref.\!\cite{Bazhanov:2003ni}, the Schr$\ddot{\rm o}$dinger equation was studied
\bea
\bigg(- {\frac{{\rd}^2}{{\rd} x^{2}}} +V(x)-E\bigg)\, \psi=0
\nonumber
\eea
with the so-called Monster potentials of the form
\bea
V(x)=\frac{\ell(\ell+1)}{x^2}+x^{2\alpha}-2\, \frac{{\rd}^2}{{\rd} x^2}\sum_{j=1}^N\log\big(x^{2\alpha+2}
-\tfrac{\alpha+1}{\alpha}\, w_j\big)\ .
\nonumber
\eea
Here the set of complex numbers ${\boldsymbol w}\equiv \{w_j\}_{j=1}^N$ satisfy the
 system of $N$ algebraic equations
\bea\label{jssysys}
\sum_{m=1\atop m\not= j}^N
\frac{w_j \big(w_j^2+(3+\alpha)(1+2\alpha) w_j w_m+\alpha (1+2\alpha) w_m^2\big)}{(w_j-w_m)^3}-
\frac{w_j}{4}+\frac{(2\ell+1)^2-4\alpha^2}{16(\alpha+1)}=0\, .
\eea
With these constraints imposed on the positions of the singularities 
 any solution of the Schr$\ddot{\rm o}$dinger equation 
 is monodromy free  everywhere  except for $x=0$ and $x=\infty$ for any value of $E$.
In other words the solutions remain single-valued 
in the vicinity of each singularity specified by $w_j$.
For this reason, these
 points are referred to as  apparent  singularities.
\bigskip

Assuming that $\alpha>0$, one can consider the standard spectral problem for
the ODE defined on the  ray $x>0$.  This leads to a discrete spectral set 
$\{E_j\}_{j=1}^\infty$. It was explained in Appendix A of ref.\!\!\cite{Bazhanov:2003ni} 
that this set can be obtained through
the solution of the exact Bohr-Sommerfeld quantization condition, which takes the form
 \bea\label{jaasusaee}
 \frac{1}{2\pi\ri}\ \log\bigg[\frac{D(q^{+2} E_j)}{D(q^{-2}E_j)}\bigg]=\frac{1}{2}-\frac{\ell+
 \frac{1}{2}}{\alpha+1}-
 { m}_j\ 
 \ \ \ \ \ \ (j=1,2,\ldots)\ .
 \eea
 Here $q=\re^{\frac{\ri\pi}{1+\alpha}}$, while  $D(E)$ denotes the spectral determinant
that for $\alpha>1$ is given by the convergent product
 \bea\label{issus}
D(E)=\prod_{n=1}^\infty\Big(1-\frac{E}{E_n}\Big)\ .\nonumber
\eea
For a given Monster potential the integers ${m}_j$ appearing in the r.h.s. of \eqref{jaasusaee} are fixed unambiguously once
the branch of the logarithm is specified.
 Comparison with the scaling form of the BA equation \eqref{jaasusa}
suggests that  for any low energy state the set of 
 scaling Bethe roots $\{s_j\}_{j=1}^\infty$ coincides
 with the spectral set $\{E_j\}_{j=1}^\infty$ for a certain Monster potential up to an overall factor,
 provided that the following identifications of the parameters are made
 \bea\label{sosaiisa}
 \beta=\frac{1}{\sqrt{1+\alpha}}\ ,\ \ \ \ \ \ \ \ \ \ \ \ {\tt K}=\frac{\ell+\frac{1}{2}}{\alpha+1}\ .
 \eea
Using the WKB approximation one can show that for $j\to \infty$
\bea\label{hsasyasy}
\big( E_j/\nu\big)^{\frac{1+\alpha}{2\alpha}}=j-\tfrac{1}{2}+\tfrac{1}{4}\, (2\ell+1)+o(1)\ ,
\eea
where the constant $\nu$ reads explicitly as
\bea\label{sosisisis}
\nu=
\bigg[\frac{2\sqrt{\pi}\Gamma(\frac{3}{2}+\frac{1}{2\alpha})}
{\Gamma(1+\frac{1}{2\alpha})}\bigg]^{\frac{2\alpha}{\alpha+1}}\ .
\eea
Comparing  \eqref{hsasyasy} with the asymptotic formula for $s_j$ \eqref{jssus} 
implies that
 \bea
 s_j=E_j/\nu \ .
 \eea
 The same is of course true for the set  $\{{\bar s}_j\}_{j=1}^\infty$, which will correspond to
 another Monster potential 
\bea
{\bar V}(x)=\frac{\bar{\ell}(\bar{\ell}+1)}{x^2}+x^{2\alpha}-2\, \frac{{\rd}^2}{{\rd} x^2}\sum_{j=1}^{\bar{N}}\log\big(x^{2\alpha+2}-\tfrac{\alpha+1}{\alpha}\, \bar{w}_j\big)
\nonumber
\eea
characterized by  the same value of $\alpha$, 
 but with $\bar{\ell}$ such that 
\be
\bar{{\tt K}}=\frac{\bar{\ell}+\frac{1}{2}}{\alpha+1}
\nonumber
\ee
 and
 another set %of apparent singularities
 ${\bar {\boldsymbol w}}=\{{\bar w}_j\}_{j=1}^{\bar N}$.
 \bigskip

In ref.\!\cite{Bazhanov:2003ni} it was argued that  for given $N$ and generic values
of  $(\alpha,\ell)$,   the  number of distinct  Monster 
potentials coincides with the number of integer partitions of $N$, i.e., ${\tt par}(N)$:
\bea
\sum_{N=0}^\infty {\tt par}(N)\ z^N=\prod_{m=1}^\infty(1-z^m)^{-1}=1+z+2\,z^2+3\,z^3+5\,z^4+\ldots\ .
\nonumber
\eea
This  important property was recently proven by  D. Masoero\cite{Masoero}.
We now note that the level subspace  ${\cal F}_{\tt p}^{(N)}$, which is spanned on the vectors of the form
\bea
\prod_j a_{-n_j}\,|{p}\rangle\ \ \ \ \ \ \ {\rm with}\  \ \ \ \  n_j>0\ \ \ \  {\rm and}\ \ \ \sum_j n_{j}=N\ ,
\nonumber
\eea
has dimensions ${\rm par}(N)$, i.e., it coincides with the number of distinct Monster potentials for fixed $N$.
This suggests that the scaling limit of any low energy Bethe state is described by
\be\label{oasaososa}
\lim_{L\to\infty}{\mathfrak N}^{-\frac{1}{2}}\, |\,{\Psi}_L\rangle=|\bar{\bm{w}}\rangle\otimes
|\bm{w}\rangle\ ,
\ee
where $\mathfrak{N}$ stands for the norm  \eqref{aosaiosia}.
The states $|\bm{w}\rangle$, labeled by the set %of apparent singularities
$\bm{w}=\{w_j\}_{j=1}^N$,  form the basis in the level subspace ${\cal{F}}_{ p}^{(N)}$.
Similarly $|\bar{\bm{w}}\rangle$, labeled by 
the fully independent set $\bar{\bm{w}}=\{\bar{w}_j\}_{j=1}^{\bar N}$, are a basis of
$\bar{\cal{F}}_{\bar{{ p}}}^{(\bar{N})}$. Thus we can see the emergence of
 the general structure \eqref{ososos232} through the scaling limit of the Bethe states.
%Thus we see the recover the general structure
% \eqref{ososos232}.
Also \eqref{ososos232} implies that
\bea\label{kssssi}
2p\,=\beta\,S^z
 +\beta^{-1} \,({\tt k}+{\tt w}) =
\frac{\ell+\frac{1}{2}}{\sqrt{\alpha+1}}\ ,\nonumber\\[-0.2cm]&&\\[-0.2cm]
2{\bar p}\,=\beta\,S^z
 -\beta^{-1}\, ({\tt k}+{\tt w}) =
\frac{{\bar \ell}+\frac{1}{2}}{\sqrt{\alpha+1}}\ .\nonumber
\eea
\bigskip

Of course there are many ways to introduce a basis in the level subspaces
${\cal{F}}_{{p}}^{(N)}$. The special property of the states
$|{\boldsymbol w}\rangle$
 is that they are consistent with  a certain
integrable structure that was  studied in refs.\cite{Bazhanov:1994ft,Bazhanov:1996dr,Bazhanov:1998dq}.
Here we'll present a few results from those papers.
Using the chiral Bose field $\partial\phi$ \eqref{issss}
one can construct a commuting set of local Integrals of Motion (IM)
$\{\mathbb{I}_{2n-1}\}_{n=1}^\infty$ with spin $2n-1=1,3,\ldots$\ :
\be\label{juusss}
[\mathbb{I}_{2n-1},\mathbb{I}_{2m-1}]=0\ .\nonumber
\ee
By local we mean that each $\mathbb{I}_{2n-1}$ is given by an integral over the local density 
built out of  $\partial\phi$ and its derivatives
\bea
\mathbb{I}_{2n-1}=\int_0^{2\pi}\frac{\rd x_1}{2\pi}\ T_{2n}\ .
\eea
For example, the first three local densities read explicitly as
\bea\label{sisisisi}
T_2&=&-\,(\partial\phi)^2\nonumber\\[0.2cm]
T_4&=&+\,(\partial\phi)^4+
\big(1-\rho^2 \big)
\ (\partial^2\phi)^2\nonumber\\[0.2cm]
T_6&=&-\,(\partial\phi)^6-\tfrac{1}{24}\ (19-32\rho^2+12\rho^4)\ (\partial^3\phi)^2-
5\, (2-\rho^2)\ (\partial^2\phi)^2(\partial\phi)^2
\nonumber\\[0.2cm]
&&\hskip 0cm\cdots\\
T_{2n}&=&(-1)^n\ (\partial\phi)^{2n}+\ldots \ .\nonumber
\eea
The dots  in the last line stand for the differential polynomials that involve, together with 
$\partial\phi$, also the higher derivatives  $\partial^2\phi,\,\partial^3\phi,\ldots\ $.
The set of local IM depend on a single parameter $\rho$ which can be an arbitrary
complex number  in general. If one makes the identification
\bea\label{rhoeq123}
\rho=\beta^{-1}-\beta=\frac{\alpha}{\sqrt{1+\alpha}}\ ,
\eea
then 
all the local IM are simultaneously diagonalized in the basis $|{\boldsymbol w}\rangle$,
i.e.,
\bea\label{eigeq1}
\mathbb{I}_{2n-1}\,
|{\boldsymbol w}\rangle=I_{2n-1}({\boldsymbol w})\ |{\boldsymbol w}\rangle\ .
\eea
 The eigenvalues of the first three IM  are given by eqs.(29) in ref.\!\cite{Bazhanov:2003ni}.
In particular 
\bea\label{eigeq2}
 {{ I}}_1({\boldsymbol w})&=&{ I}^{(0)}_1\big(\sqrt{p^2+N}\,\big)=
 { I}^{(0)}_1({p})+N\nonumber\\
\\[-0.5cm]
 {{ I}}_3({\boldsymbol w})&=&{ I}^{(0)}_3\big(\sqrt{p^2+N}\,\big)+
 \sum_{j=1}^N w_j\ .\nonumber
\eea
The higher spin integrals of motion
 ${{ I}}_{2n-1}({\boldsymbol w})$ turn out to be symmetric polynomials
of order $n-1$ w.r.t. the variables $\{w_j\}_{j=1}^N$. In their turn,
the level zero eigenvalues ${ I}_{2n-1}^{(0)}(p)$ are polynomials of order
$n$ w.r.t. $p^2$. For example
\bea
I_1^{(0)}(p)&=&p^2-\tfrac{1}{24}\nonumber\\[0.2cm]
I_3^{(0)}(p)&=&
p^4-\tfrac{1}{4}\ p^2+\tfrac{1}{960}\ (9-4\rho^2)\nonumber\\[0.2cm]
&&\cdots\nonumber\\
I_{2n-1}^{(0)}(p)&=&p^{2n}+O\big(p^{2n-2}\big)\ .\nonumber
\eea
Since the local IM act invariantly on each level subspace ${\cal F}_{p}^{(N)}$, 
the diagonalization problem \eqref{eigeq1} reduces to that
of finite ${\tt par}(N)\times {\tt par}(N)$ dimensional mutually commuting matrices for any given $N$. 
Let's present some explicit formulae for  $|{\boldsymbol w}\rangle$ 
for the lowest excited states.
\bigskip

For $N=1$, when the Monster potential contains only one
apparent singularity,
the equations \eqref{jssysys} dramatically simplify. Their solution is
\bea
%\frac{\alpha w_1}{\alpha+1}
w_1=(2p-\rho)(2p+\rho)\ ,\nonumber
\eea
which is expressed in terms of ${p}$ and $\rho$, related to $\ell$ and $\alpha$ as in
eqs.\,\eqref{kssssi},\,\eqref{rhoeq123}.
Since ${\dim {\cal F}}_{p}^{(1)}=1$ one has that
\bea\label{sjssusu}
\big|{ \boldsymbol w}^{(1)}\big\rangle=\big[{F}^{(1)}\big]^{-\frac{1}{2}}\ \ a_{-1}\, |\,{p}\rangle\ .
\eea
Here ${F}^{(1)}$ is the overall normalization constant, which remains undetermined.
\bigskip

For $N=2$ there are two solutions of eq.\eqref{jssysys}, which will be denoted by
${\boldsymbol w}^{(2,+)}=(w^{+}_1,w^{+}_2)$  and  ${\boldsymbol w}^{(2,-)}=(w^{-}_1,w^{-}_2)$.
Explicitly one can show that
%\bea
%w_1^{(\pm)}&=&\frac{2\, \omega_\pm (\omega_\pm+\alpha+1)(\omega^2_\pm
%-\alpha(2+\alpha))}{\alpha(\alpha+1)}\\[0.2cm]
%w_2^{(\pm)}&=&\frac{2 \omega_\pm 
%(\omega_\pm-\alpha-1)(\omega^2_\pm-\alpha(2+\alpha))}{\alpha(\alpha+1)}\ ,\nonumber
%\eea
\bea
%\frac{\alpha w_1^{(\pm)}}{\alpha+1}
w_1^{\pm}&=&2\, \omega_\pm \big(\omega_\pm+\beta^{-1}\big)\big(\omega^2_\pm
+\beta^2-\beta^{-2}\big)\nonumber\\[0.2cm]
%\frac{\alpha w_2^{(\pm)}}{\alpha+1}
w_2^{\pm}&=&
2\, \omega_\pm \big(\omega_\pm-\beta^{-1}\big)\big(\omega^2_\pm
+\beta^2-\beta^{-2}\big)
\ ,\nonumber
\eea
where
\bea
\omega_\pm
%&=&\frac{1}{2}\ \sqrt{(\alpha+3)(2\alpha+1)\pm
%\sqrt{(\alpha-1)^2(2\alpha+1)^2+32 (\alpha+1)^2\,{p}^2\!\!\phantom{\big|}}}\nonumber\\
&=&\frac{1}{2}\ \sqrt{(1+2\beta^2)(2\beta^{-2}-1)\pm B}\nonumber
\eea
with
\bea
B=\sqrt{(2\rho^2-1)^2+32 p^2}\ >0\ .\nonumber
\eea
The corresponding  basis states $|{\boldsymbol w}^{(2,\pm)}\rangle\in {\cal F}_{p}^{(2)}$ are given by
\bea\label{aisaiisiass}
\big|{\boldsymbol w}^{(2,\pm)}\big\rangle=\Big[ {F}^{(2,\pm)}\Big]^{-\frac{1}{2}}\ \bigg(\,
a_{-1}^2-\frac{4p}{1-2\rho^2\mp B}\ a_{-2}\,\bigg)\, |{ p}\rangle\ .
\eea
%with
%\bea
%c_\pm&=&\frac{4\,(1+\alpha)\, {p}}{(1-\alpha)(1+2\alpha)\mp
%\sqrt{(1-\alpha)^2(1+2\alpha)^2+32\, (1+\alpha)^2{ p}^2\!\!\phantom{\big|}}}\ .
%\\
%&=&\frac{(1+\alpha)\, p}{1+2\alpha-\omega_\pm^2}
%\eea

 \section{\label{sec6}Scaling limit of the norms}

The states $|{\boldsymbol  w}\rangle$ that appear in the scaling limit
 \eqref{oasaososa} diagonalize the full set of local IM \eqref{eigeq1}.
It is expected that for generic values of $(\beta,p)$, this
 allows one to specify the states $|{\boldsymbol  w}\rangle$ up to an overall normalization.
To resolve this last ambiguity, we should discuss the natural Hermitian structure
appearing in the spin chain and its scaling limit.
\bigskip

The space of states of the spin $\half$ chain of length $L$ is the
tensor product of $L$ copies of the two-dimensional  complex vector space. 
The positive definite  inner product for this space is induced by that 
of each two-dimensional component. The latter is defined as 
$\langle\sigma|\sigma'\rangle=\delta_{\sigma,\sigma'}$, where 
$|\pm \rangle$ stands for the two basis vectors such that $\sigma^z|\pm\rangle=\pm|\pm\rangle$. 
For generic values of the twist parameter  ${\tt k}\not=0$, any two Bethe states corresponding to different
solutions of the BA equations  turn out to be orthogonal w.r.t. this
inner product.
We'll return to this important property later, in sec.\,\ref{sec10}.
\bigskip

On the other hand, 
the  space  $\oplus_{{\bar p},p}\,{\bar {\cal F}}_{\bar p}\otimes {\cal F}_{ p}$ admits a natural
positive definite inner product specified unambiguously by the conjugation
condition for the Heisenberg generators,
\bea\label{aisisaias}
a_{m}^\dagger=a_{-m}\ ,\ \  \ \ \ \ \ \ \ {\bar a}_{m}^\dagger={\bar a}_{-m}\ \ \  \ (\forall m)
\eea
 together with the relations for the highest vector
 \bea\label{aosia}
 \langle p\,|\, p'\rangle=\delta_{p,p'}\ ,\ \ \ \ \ \ \langle{\bar p}\,|\,{\bar  p}'\rangle=\delta_{{\bar p},{\bar p}'}\ .
 %\nonumber
 \eea
 It is important that this Hermitian structure is consistent with the integrable structure described in the previous section.
 In particular, all the local IM are Hermitian operators w.r.t. the conjugation \eqref{aisisaias},
 \bea
 {\mathbb I}_{2n-1}^\dagger={\mathbb I}_{2n-1}
 %\nonumber
 \eea
and therefore, for generic values of $p$, one may  expect that the states $|{\boldsymbol w}\rangle$ and $|{\boldsymbol w}'\rangle$ 
 corresponding to different sets ${\boldsymbol w}$ and ${\boldsymbol w}'$ to be orthogonal.
Thus, we come to the conclusion that the 
 natural Hermitian structure in the  $XXZ$ spin chain of finite length becomes  
 that defined by formulae \eqref{aisisaias} and \eqref{aosia} in the scaling limit.
 Then eq.\eqref{oasaososa} implies that the states $|{\boldsymbol w}\rangle$ form  an
 orthonormal basis in ${\cal F}_{p}$:
 \bea\label{isisisis}
 \langle {\boldsymbol w}\,|\, {\boldsymbol w}'\rangle=\delta_{{\boldsymbol w},{\boldsymbol w}'}
 \eea
 and similarly for $|{\bar {\boldsymbol w}}\rangle$.
 The last condition fixes the overall normalization of $|{\boldsymbol w}\rangle$ up to
 an inessential sign factor.\footnote{
The  $\cal{C}\cal{P}\cal{T}$ transformation acts on the Bethe wave function as 
$ \Psi(x_1,\ldots, x_M)\mapsto
\Psi^*(L+1-x_M,\ldots, L+1-x_1)$. As it follows
from the formulae \eqref{aisaiasu},\,\eqref{aisasau},\,\eqref{asisaisa},\,\eqref{kaajay} and \eqref{asdi1231}
under the $\cal{C}\cal{P}\cal{T}$ transformation the Bethe state gains an overall phase factor
$\re^{\ri\vartheta}$
with $\vartheta=(L+1)\sum_m p_m$.
By a simple modification of the
Bethe wave function, the Bethe state can be adjusted so
 that ${\cal \hat{C}\hat{P}\hat{T}}\,|\Psi_L\rangle=|\Psi_L\rangle$.
Together with eq.\,\eqref{oasaososa}, this implies that
${\cal \hat{C}\hat{P}\hat{T}}\,|\bm{w}\rangle=|\bm{w}\rangle$ and
${\cal \hat{C}\hat{P}\hat{T}}\,|\bar{\bm{w}}\rangle=|\bar{\bm{w}}\rangle$.
Since the $\cal{C}\cal{P}\cal{T}$ transformation acts in the Fock space as
$c\,\prod_j a_{-m_j}|p\rangle \mapsto c^*\,\prod_j a_{-m_j}|p\rangle$,
where $c$ is a c-number, this allows one to resolve the phase ambiguity in the
normalization of $|\bm{w}\rangle$ and $|\bar{\bm{w}}\rangle$.} 
For example, the undetermined constants  in 
 eqs.\eqref{sjssusu} and \eqref{aisaiisiass} read as follows:
 \bea
 {F}^{(1)}=\frac{1}{2}\ ,\ \ \ \ \ \  \ {F}^{(2,\pm)}=
 \frac{B}{B\pm (2\rho^2-1)}\ .
% \frac{\omega_\pm^2-\frac{1}{4}\, (1+2\beta^2)(2\beta^{-2}-1)}{\omega_\pm^2+\beta^{2}-2}\ .
 \eea
% with $B=\sqrt{(2\rho^2-1)^2+32 p^2}$.
 
 \bigskip
 Let us now turns to the norms ${\mathfrak N}$ appearing in eq.\,\eqref{oasaososa}. Our numerical work led us to the following
 
 \medskip
 {\bf Conjecture I:}  The scaling exponent in the asymptotic formula \eqref{aososaik}
 is given by
 \bea\label{ossiss}
 \eta=\tfrac{1}{6}-\tfrac{1}{3}\,\rho^2
 -2\rho\beta S^z
 -4 \,I_1({\boldsymbol w})-4 \,{\bar I}_1({\bar {\boldsymbol w}})\ ,
 \eea
 while the amplitude  ${\mathfrak  N}_\infty$ factorizes as
 \bea\label{scalingN1}
 {\mathfrak  N}_\infty={\cal N}_{\bar p}({\bar {\boldsymbol w}})\, {\cal N}_p({\boldsymbol w})\ .
 \eea
%\bigskip
%  Notice that with the above conjecture, eq.\eqref{oasaososa} can be rewritten as
 %\be\label{oasaososa11a}
%\lim_{L\to\infty}{L}^{-\frac{\eta }{2}}\, \re^{-\frac{1}{2}({\cal A}_2L^2+{\cal  A}_1 L)}\, |\,{\Psi}_L\rangle= 
%|\bar{\bm{\psi}}\rangle\otimes
%\bm{\psi}\rangle
%\ee
%with
%\bea
%|{\bm \psi}\rangle=\sqrt{{\cal N}_{p}({\boldsymbol w})}\ |\bm{w}\rangle\ ,\ \ \ \  \  |{\bar {\bm \psi}}\rangle=
%\sqrt{{ {\cal N}}_{\bar p}({\bar {\boldsymbol w}})}\ |\bar{\bm{w}}\rangle\  .
%\eea

\section{Characterizing the Bethe states when $L$ is finite}

The scaling amplitudes ${\cal N}_{ p}$ are functions of the set $\bm{w}$, which
label the states in the Fock space.
Thus, for their numerical computation
using eqs.\eqref{aososaik},\eqref{aisiasjjs},\eqref{asdasd1122},\eqref{ossiss},\eqref{scalingN1},  
we run into an important practical problem:  
having at hand the Bethe roots corresponding to the
 Bethe state $|\Psi_L\rangle$ for a few values of $L$, how to identify
its scaling limit, i.e., to extract the sets $\bm{w}$ and $\bar{\bm{w}}$ characterizing
 the r.h.s. of eq.\,\eqref{oasaososa}. 
\bigskip

The most straightforward approach to the problem is to study the 
energy ${\cal E}(L)$ corresponding to $|\Psi_L\rangle$. In terms of the
Bethe roots, it is given by the following expression
\be
{\cal E}=-\tfrac{1}{2}\, (q+q^{-1})\, L+\sum_{m=1}^M\frac{2\, (q-q^{-1})^2}{
\zeta_m+\zeta_m^{-1}+
q+q^{-1}}\ .\nonumber
\ee
For the critical spin chain, the large-$L$ behavior obeys the asymptotic
\be\label{oasidsid1a}
{\cal E}(L)\,\asymp\, e_\infty\,L+\frac{2\pi v_{\rm F}}{L}\,\big(\,I_1(\bm{w})+I_1(\bar{\bm{w}})+\delta_L\,\big)\ \ \ 
\ \ \ \ {\rm with}\ \ \ 
\delta_L=o(1)\ .
\ee
Here $e_\infty$ is the specific bulk energy, which in the case at hand is given by
\be
e_{\infty}=-\cos(\pi\beta^2)-\frac{2 v_{\rm F}}{\pi }\ 
\int_0^\infty\rd t\  \frac{\sinh\big(\frac{\beta t}{\rho}\big)}
{\sinh(\frac{t}{\beta\rho})\,\cosh(t)}\ ,\nonumber
\ee
while $v_{\rm F}$ is the Fermi velocity (to be compared with eq.\,\eqref{asdasd1122}):
\be
v_{\rm F}=\frac{2\sin(\pi \beta^2)}{1-\beta^2}\ .\nonumber
\ee
In view of eq.\eqref{eigeq2}, the formula \eqref{oasidsid1a} 
allows one to extract the integer 
$N+\bar{N}$  from the numerical data at finite $L$. However, it is of course insufficient to determine 
the sets $\bm{w}$ and $\bar{\bm{w}}$ characterizing the scaling limit of $|\Psi_L\rangle$.
More information can be obtained from the study of the finite size corrections to the energy,
denoted by $\delta_L$ in eq.\,\eqref{oasidsid1a}. The leading behavior of $\delta_L$ was found
in ref.\!\cite{Lukyanov:1997wq} (see also \cite{Sirker:2005ab}). For $0<\beta^2<\frac{2}{3}$, it is expressed in terms of the local integrals
of motion
\be\label{coreq1a}
\delta_L=-\bigg(\frac{2\pi}{L}\bigg)^2\,\Big(\lambda_+\,I_1\,\bar{I}_1+\lambda_-\,(I_3+\bar{I}_3)\Big)+o(L^{-2})\ ,
\ee
where the constants $\lambda_\pm$ are known explicitly and are presented in that work.
For the case when $\frac{2}{3}<\beta^2<1$, the corrections take the form
\be\label{coreq1b}
\delta_L=-\tfrac{2}{\pi}\,\beta^8\,\sin\big(\tfrac{2\pi}{\beta^2}\big)\,\Bigg[\frac{\sqrt{\pi}\,\Gamma\big(1+\frac{\beta}{2\rho }\big)}{L\,
\Gamma\big(1+\frac{1}{2\beta\rho}\big)}\Bigg]^{\frac{4\rho }{\beta}}\ \tilde{H}_1\,\bar{\tilde{H}}_1+
o\big(L^{-\frac{4\rho}{\beta}}\big)\ .
\ee
Here  $\tilde{H}_1$ and $\bar{\tilde{H}}_1$ are the eigenvalues of the so-called dual non-local integrals of motion 
$\tilde{\mathbb{H}}_1$ and $\bar{\tilde{\mathbb{H}}}_1$ respectively\ \cite{Bazhanov:1996dr}. Contrary to the local IM,
it is rather tedious to compute
the eigenvalues of these operators
for the excited states. 

\begin{figure}
\centering
\begin{tikzpicture}
%\node at (-4.5,3.4) {$\Pi(L)$};
\node at (5.4,-2.65) {$L$};
\node at (0,0) {\includegraphics[width=10cm]{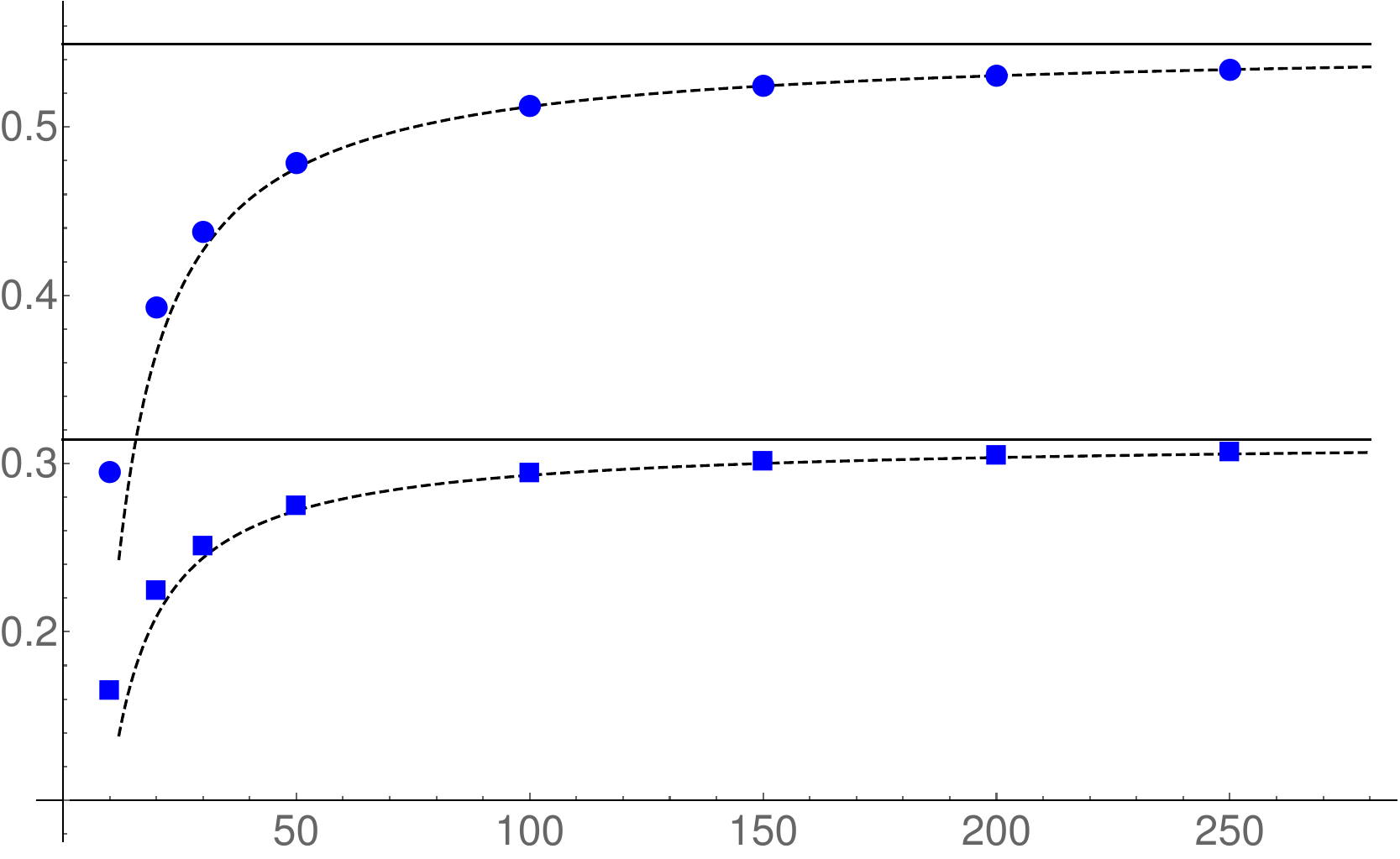}};
\node at (4.5,3.13) {\small $\big[R^{(2,-)}_p\big]^2$};
\node at (4.4,0.3) {\small $\big[R^{(0)}_{\bar p}\big]^2$};
\end{tikzpicture}
\caption{\small \!\!\!\! The Bethe roots for the family of states $|\Psi_L\rangle$ with
$N=2$, $\bar{N}=0$ and for increasing $L=10,20,30,\ldots,250$
have been used to compute the l.h.s. of eqs.\,\eqref{prodeqA} and \eqref{prodeqB}.
These numerical values  have been rescaled by the factors
$(\frac{L}{\pi})^{4\rho p}\,\big(4(1-\beta^2)\big)^{-L}$ for the l.h.s. of \eqref{prodeqA} and
$(\frac{L}{\pi})^{4\rho \bar{p}}\,\big(4(1-\beta^2)\big)^{-L}$ for that of \eqref{prodeqB},
 and the result is depicted in the 
plot by the circles  and squares respectively.
The dashed lines represent the curves $0.548827-3.66513/L$ and
$0.314091-2.10573/L$, which were obtained by fitting the data.
The limiting values predicted from eqs.\,\eqref{prodeqA}\,\eqref{prodeqB}, $\big[R^{(2,-)}_p\big]^2=0.549236$ 
and $\big[R^{(0)}_{\bar p}\big]^2=0.314304$,
are given by the solid lines.
The parameters were taken to be $S^z=0$, $\beta^2=\frac{2}{5}$
and $\pi{\tt k}=-\frac{3}{20}$.\label{fig4a}}
\end{figure}
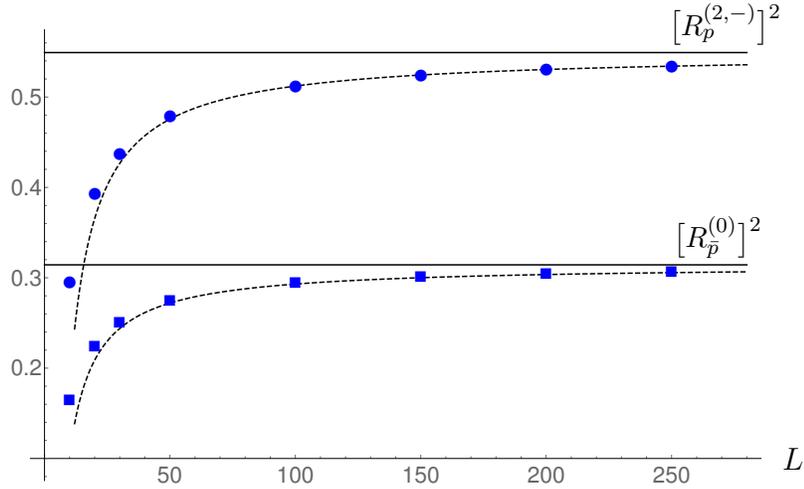

\bigskip
The formulae \eqref{coreq1a},\,\eqref{coreq1b} 
are sometimes useful for finding $\bm{w}$ and $\bar{\bm{w}}$, especially when  $0<\beta^2<\frac{2}{3}$,
for which the leading behavior of $\delta_L$ is expressed in terms of the local IM. However,
there exists  a more effective method for identifying the scaling limit of the Bethe state.
To explain it, we first make the following remark. 
As was mentioned before, the functions $A(s)$ and $\bar{A}(s)$ defined in eq.\,\eqref{issus1a}
can be regarded as the scaling limit of the eigenvalues of the lattice $Q$-operator \eqref{oasidoi192}.
It turns out that it is possible to construct the operator $\mathbb{A}(s)$ acting in the Fock space
${\cal F}_p$ whose eigenvalue on the state $|\bm{w}\rangle$ coincides with $A(s)$. 
For the explicit formulae, we refer the reader to the original papers \cite{Bazhanov:1996dr,Bazhanov:1998dq}.
Here we just mention that the sets of
mutually commuting local $\{\mathbb{I}_{2n-1}\}_{n=1}^\infty$ 
and dual non-local $\{\tilde{\mathbb{H}}_n\}_{n=1}^\infty$
IM are generated through the large-$s$ operator-valued asymptotic expansion of $\mathbb{A}$.
For $s\to \infty$ and with $|\arg(-s)|<\pi$, the  expansion reads as follows \cite{Bazhanov:1996dr,Bazhanov:2003ni}
\bea\label{Apeq1}
 {\mathbb  A}(s)
 &\asymp& {\mathbb R}\ \ (-s)^{-\frac{p}{\beta}}\ \  \exp\Big( \tfrac{\pi}{\sin(\frac{\pi }{2\beta\rho})}\, 
(-s)^{\frac{1}{2\beta\rho}}\Big) 
\\ 
%\frac{\pi\, C}{\cos(\frac{\pi}{n})}\ 
&\times&\exp\bigg(\, \sum_{n=1}^\infty  b_{n}\ { {\mathbb  I}}_{2n-1}\ (-\nu s)^{\frac{1-2n}{2\beta\rho}}
%+\big(\ri s-\tfrac{n p}{n+2}\big)\ \tfrac{n+2}{2n}\ \log\big(-\zeta^2\big)
%+{\mathbb C}
+
\sum_{n=1}^\infty\,  c_n\,{\tilde {\mathbb H}}_n\
 (-\nu s)^{- \frac{n}{\beta^2}} \, \bigg)\ .\nonumber
 \eea
 Here the constant $\nu$ is given by eq.\eqref{sosisisis},
%\bea
%\nu=
%\Bigg[\frac{2\sqrt{\pi}\Gamma(\frac{3}{2}+\frac{\beta}{2\rho})}
%{\Gamma(1+\frac{\beta}{2\rho})}\Bigg]^{2\beta\rho}\ .
%\eea 
while the numerical coefficients $b_n$ and $c_n$ are not essential for our purposes
 and can be found in  sec.\,4 of ref.\!\cite{Bazhanov:2003ni}.
 The subject of our interest is the operator ${\mathbb R}$, which will be referred to as the reflection operator.
 It commutes with all the IM
 and can be diagonalized simultaneously with them:
 \bea\label{hssysys}
 {\mathbb R}\, |{\boldsymbol w}\rangle=R_{p}({\boldsymbol w})\ |{\boldsymbol w}\rangle\ .
 \eea
It is possible to show that the eigenvalue of this operator is related to the Bethe roots 
for the Bethe state $|\Psi_L\rangle$ as
 \bea\label{prodeqA}
\prod_{m=1}^M\big(\zeta_m^{-1}+q\big)\big(\zeta_m^{-1}+q^{-1}\big)&\asymp&\big (R_{ p}({\boldsymbol w})\big)^2\  \ 
\big(\tfrac{L}{\pi}\big)^{-4\rho {p}}\ 
\big(4(1-\beta^2)\big)^{L}\,\big(1+o(1)\big)\\[0.1cm]\label{prodeqB}
\prod_{m=1}^M\big(\zeta_m+q\big)\big(\zeta_m+q^{-1}\big)&\asymp& \big (R_{\bar {p}}({\bar {\boldsymbol w}})\big)^2\ \  
\big(\tfrac{L}{\pi}\big)^{-4\rho{\bar { p}}}\ 
\big(4(1-\beta^2)\big)^{L}\,\big(1+o(1)\big)\ .
\eea
Thus, the eigenvalues $R_{ p}({\boldsymbol w})$ and
$R_{\bar {p}}({\bar {\boldsymbol w}})$ can be extracted numerically 
from the Bethe roots for sufficiently large $L$ (see fig.\,\ref{fig4a}).
On the other hand, as will be discussed in the next section, there  exists a straightforward procedure
to calculate the spectrum of the reflection operator in the Fock space ${\cal F}_p$.
We found that this was the most effective way of identifying the scaling limit of the
Bethe state $|\Psi_L\rangle$.
\bigskip

For $N=0,1,2$ the eigenvalues of $\mathbb{R}$ read as follows.
For the highest state in the Fock space, $R^{(0)}$ is given by \cite{Bazhanov:1996dr,Bazhanov:1998wj}
 \bea\label{Reig1a}
 R^{(0)}_p=\beta^{1+4p\beta}\ \ 
 \frac{\Gamma(1+\frac{2p}{\beta})}{\Gamma(1+2p\beta)}\ 
 \Bigg[\frac{\Gamma(1+\frac{\beta}{2\rho})}{\sqrt{\pi}\Gamma(\frac{3}{2}+\frac{\beta}{2\rho})}
\Bigg]^{2p\rho }\ 
 \ .
 \eea
For the first level state \eqref{sjssusu}, the eigenvalue is
\bea\label{Reig1b}
 R^{(1)}_p=R^{(0)}_p\ \frac{2p+\rho}{2p-\rho}\ .
\eea
One has that for the two states at the second level, defined by eq.\,\eqref{aisaiisiass}, the corresponding eigenvalues are
\bea\label{Reig1c}
R_p^{(2,\pm)}=R_p^{(0)}\ 
\frac{\big(2p\,(2\rho^2+1)\pm\rho\,B\big)\,\big(2p\,(\beta+2\beta^{-1})\pm B \big)\,\big(2p\,(2\beta+\beta^{-1})\mp B\big)}
           {(2\rho^2-1)^2\,(2p-\rho)\,(2p+\beta-2\beta^{-1})\,(2p+2\beta-\beta^{-1})}\ ,
\eea
where $B=\sqrt{(2\rho^2-1)^2+32p^2}$.
%\bea\label{Reig1c}
 %R^{(2,\pm)}_p=R^{(0)}_p\ \frac{8p^3-2 (1+3\rho^2)\,p\mp \rho\, \sqrt{(2\rho^2-1)^2+32 p^2}}{(2p-\rho)(2p+\beta-2\beta^{-1})
 %(2p+2\beta-\beta^{-1})}\ .
 %\eea

 \section{\label{sec8}The reflection operator}

 The reflection operator is closely related to the reflection $S$-matrix of the
 Liouville CFT. The latter was discussed in detail in the seminal work of 
the  Zamolodchikov brothers \cite{Zamolodchikov:1995aa}.  Here, following this paper, we'll briefly describe its construction.
\bigskip

 Consider the Liouville CFT associated with the Lagrangian
 \bea\label{sosososd}
{\cal L}=\frac{1}{4\pi}\ (\partial_a \phi)^2+\mu\ \re^{2b\phi}\,,
\eea
where the space-time is the 2D  Euclidean cylinder equipped with the complex coordinates $z=x_2+\ri \,x_1$ and
${\bar z}=x_2-\ri \,x_1$.
Because of the scale invariance, we can assume that $x_1\sim x_1+2\pi$.
Contrary to the Gaussian model \eqref{ososos} the field $\partial\phi$ is not holomorphic for
the Liouville CFT.  However the component $T\equiv T_{zz}$ of the energy-momentum tensor, 
\bea\label{oasoos}
T=-(\partial\phi)^2+Q\ \partial^2\phi\ \ \ \ \ \ \ \ \big(Q=b+b^{-1}\big)\ ,
%\nonumber
\eea
 satisfies  the Cauchy-Riemann equation ${\bar \partial}T=0$. Being  the periodic field 
 at a given Euclidean time slice, $T$ can be expanded in the Fourier series
 \bea\label{Texpansion1}
 T(z)=-\frac{c}{24}+\sum_{n=-\infty}^\infty L_n\ \re^{-nz}\ .
% \nonumber
 \eea 
The expansion coefficients satisfy the Virasoro algebra ${(\it Vir})$  commutation relations
\bea\label{sosososoy}
[L_n,L_m]=(n-m)\, L_{n+m}+\tfrac{c}{12}\ n(n^2-1)\, \delta_{n+m,0}
%\nonumber
\eea
 with the central charge
 \bea\label{iauaua}
 c=1+6\,Q^2\ .
 \eea
Similar relations hold true for the antiholomorphic component $T_{\bar{z}\bar{z}}(\bar{z})$ of the energy momentum
 tensor.
 The space of states of the Liouville CFT  is classified using the highest weight representations of 
 $\overline{Vir}\otimes
 { Vir}$.
 The chiral part of this  Hilbert space is built from the Verma module containing the primary state
 $|v_p\rangle$:
 \bea
 L_n \, |v_p\rangle=0\ \ \ \ (n>0)\ ;\ \ \ \ L_0 \, |v_p\rangle=\big(p^2+\tfrac{Q^2}{4}\big) \, |v_p\rangle\ .
 \nonumber
 \eea
 
 \bigskip
 Let us introduce the ``zero-mode'' of the Liouville field $\phi(x)$
 \bea
 \phi_0=\int_0^{2\pi}\frac{{\rd x}_1}{2\pi}\ \phi(x)
 \nonumber
 \eea
 and consider the asymptotic domain in the configuration space where $\phi_0\to-\infty$. Then
 the exponential interaction in \eqref{sosososd} is negligible, so that $\phi(x)$ becomes a free massless field
 that can be expanded in terms of the free field oscillators as in \eqref{issss},\eqref{sisisis}
 \bea\label{issssdd}
\phi=\phi_0-\ri\, {\hat p}\ (z+{\bar z})+
\ri\, \sum_{m\not=0}\, \Big(\frac{ a_m}{m}\ \re^{-m z}+\frac{ \bar{a}_m}{m}\ \re^{-m \bar{z}}\Big)\ .
\nonumber
\eea
Here ${\hat  p}=\frac{1}{2\ri}\ \frac{\partial}{\partial \phi_0}$ is the momentum conjugate to the 
zero-mode. In the asymptotic domain, 
the Virasoro generators are represented as follows
\bea\label{ososossi}
L_n&=&\sum_{m\not=0,n}a_m a_{n-m}+\big(2{\hat  p}+\ri Q\, n\big)\, a_n\ \ \ \  \ \ \ \ \ (n\not=0)\nonumber\\
L_0&=&2\sum_{m>0}a_{-m} a_{m}+{\hat p}^2+\tfrac{Q^2}{4}\ .
\eea
The above relations suggest  that the Fock space ${ \bar {\cal F}}_p\otimes{ {\cal F}}_{p}$
for  $p={\bar p}>0$ can be interpreted as the space of ``in''-asymptotic states of the Liouville
CFT, while ${\bar {\cal F}}_p\otimes{ {\cal F}}_{p}$ with $p={\bar p}<0$ are identified
with the ``out''-states (see ref.\!\cite{Zamolodchikov:1995aa} for a further discussion). In such a situation it is
natural to introduce  the reflection $S$-matrix intertwining the spaces of in- and out- asymptotic states:
\bea
{\hat S}_{\rm L}\ :\ \ \ {\bar {\cal F}}_p\otimes{ {\cal F}}_{p}
\mapsto \ {\bar {\cal F}}_{-p}\otimes{ {\cal F}}_{-p}\ \  \ \ \ \ 
(p>0)\ .
\nonumber
\eea
It is readily established that
\bea
{\hat S}_{\rm L}=S^{(0)}(p)\ {\hat {\bar s}}(p)\otimes {\hat { s}}(p)\ ,
\nonumber
\eea
where $S^{(0)}(p)$ is a certain phase factor
while ${\hat {\bar s}}(p)$ and  ${\hat { s}}(p)$ are properly normalized
operators acting in the  chiral Fock spaces.  In particular
\bea
{\hat { s}}(p)\ :\ \ \ {\cal F}_p\mapsto {\cal F}_{-p}\ ,\ \ \ \ \ \ \  {\hat { s}}(p)\,|p\rangle=|-p\rangle\ .
\nonumber
\eea
\smallskip

The action of the operator ${\hat { s}}(p)$ is fully determined by the conformal symmetry and
is constructed in the following way.
One can consider the oscillator basis in the Fock space ${\cal F}_p$
formed by the vectors 
\bea\label{eopwop12}
{\boldsymbol a}_{I}(p)=a_{-i_m}\cdots a_{-i_1}\ |p\rangle\,,
\nonumber
\eea
where $I$ stands for the multi-index $I=(i_1,\cdots, i_m)$ with $1\leq i_1\leq i_2\ldots\leq i_m$.
On the other hand,  formulae \eqref{ososossi} define the structure of the Virasoro highest weight representation
on the Fock space  with  $|{v}_p\rangle=|p\rangle$.
There is a natural basis in ${\cal F}_p$  that is associated with this structure
\bea
{\boldsymbol L}_{I}=L_{-i_m}\cdots L_{-i_1}\ |v_p\rangle\  \ \ \ \ \ \ \ (1\leq i_1\leq i_2\ldots\leq i_m)\ ,
\nonumber
\eea
where again $I=(i_1,\cdots, i_m)$.
The two bases are of course linearly related, so that
\bea
{\boldsymbol L}_{J}={\boldsymbol a}_I(p)\ {\Omega^I}_J(p)\ .
\nonumber
\eea
The matrix elements of the operator ${\hat { s}}(p)$ are given by
\bea\label{ossisisi}
 {\big[s(p)\big]^J}_I= {\Omega^J}_A(-p){\big[\Omega^{-1}\big]^A}_I(p)\ :\ \ \ 
{\hat { s}}(p)\ {\boldsymbol a}_{I}(p)={\boldsymbol a}_{J}(p)\ {\big[s(p)\big]^J}_I\ .
%\nonumber
\eea
\medskip

It should be emphasized that, since this operator intertwines
different Fock spaces, the problem of its diagonalization does not make sense.
However one can introduce another intertwiner, the ``$C$-conjugation'',  whose action is defined by the condition
\bea\label{oosostr}
{\hat  C}\ :\ \ \ \ {\hat C}\, a_{n}\, {\hat C}=-a_{n}\ ,\  \ \ \ {\hat C}\,|p\rangle=|-p\rangle\,,
%\nonumber
\eea
so that ${\hat  C}{\hat { s}}(p)$  acts invariantly in the Fock space
\bea\label{owieoi223}
{\hat  C}{\hat { s}}(p)\, :\ \ \ \  {\cal F}_{ p}\mapsto  {\cal F}_{ p}\ \ \ \ \ \ {\rm and} \ \ \ \ \ \   {\hat  C}{\hat { s}}(p)\,|p\rangle=|p\rangle\ .
\eea
In ref.\!\!\cite{Zamolodchikov:1995aa}  it was pointed out that this operator commutes with the action of the local IM defined by
eqs.\eqref{juusss},\,\eqref{sisisisi} with  the parameter $\rho$ substituted by $-\ri Q$.
In connection to this, we note that  the local IM   can be re-written  in terms of the field  $T(z)$ \eqref{Texpansion1} 
and its derivatives only:
\bea\label{sisoisss}
{\mathbb I}_1=\int_0^{2\pi}\frac{\rd x_1}{2\pi}\ T\ ,\ \ \
{\mathbb I}_3=\int_0^{2\pi}\frac{\rd x_1}{2\pi}\ T^2
\ ,\ \ \ \ {\mathbb I}_5=\int_0^{2\pi}\frac{\rd x_1}{2\pi}\ \big(\, T^3-\tfrac{c+2}{12}\, (\partial T)^2\, \big)\ ,
\eea
where $c$ is the central charge  \eqref{iauaua} of the Virasoro algebra \eqref{sosososoy}.
In general, one has
\bea\label{gsfss}
{\mathbb I}_{2n-1}=\int_0^{2\pi}\frac{\rd x_1}{2\pi}\, \big(\,T^n+\ldots\,\big)\ ,
\eea
where the dots stand for the  terms involving $\partial^m T\ (m>0)$. The limit $c\to\infty$ can be understood as
a certain classical limit  such that  $\{{\mathbb I}_{2n-1}\}_{n=1}^\infty$ becomes the set of IM for the classical KdV equation
\!\cite{Eguchi:1989hs,Kupershmidt:1989bf,Bazhanov:1994ft}.
Thus, the operator ${\hat  C}{\hat s}$ \eqref{owieoi223} is part of the quantum  
KdV integrable structure studied in refs.\!\cite{Bazhanov:1994ft,Bazhanov:1996dr,Bazhanov:1998dq}.
It should be pointed out that the Liouville reflection $S$-matrix itself is defined by  the conformal
symmetry alone and does not assume the presence of any integrable structure.
\bigskip

The reflection operator appearing in the previous section 
is given by
\bea\label{sosoisis}
{\mathbb R}=R^{(0)}_p\ \big[{\hat  C}{\hat s}(p)\big]^{-1}\,,
\eea
where the parameters are identified as follows
\bea
Q=\ri\rho\ ,\ \ \  \ b=-\ri\beta\ .
\nonumber
\eea
Since all the matrix elements ${\big[s(p)\big]^J}_I$ are rational functions of $b$, the operator $\mathbb{R}$ 
commutes with the local IM in the domain where $0<\beta\le 1$. 
A useful relation, that follows immediately from eqs.\,\eqref{ossisisi},\eqref{oosostr},\eqref{sosoisis} and \eqref{Reig1a},
is that 
\bea
R_p({\boldsymbol w}\big)R_{-p}({\boldsymbol w}\big)=R_p^{(0)}R_{-p}^{(0)}=\frac{\sin(2\pi p\beta)}{\sin(2\pi p\beta^{-1})}
\eea
for any set ${\boldsymbol w}$.
\bigskip

Formula \eqref{sosoisis} provides an effective  tool for the calculation of
the spectrum of the reflection operator.
The eigenvalues  for the first few levels are given by 
eqs.\,\eqref{Reig1a}-\eqref{Reig1c}.
For the higher levels $N\ge 3$, the 
eigenvalues of ${\mathbb R}$ turn out to be rather cumbersome.
However their product for a given level $N$, i.e., ${\rm det}_N({\mathbb R})$, admits a remarkably simple structure
\bea
{\rm det}_N({\mathbb R})=\prod_{\boldsymbol w^{(N)}}R_p({\boldsymbol w}^{(N)}\big)=\big[R_p^{(0)}\big]^{{\tt par}(N)}\ 
\prod_{1\leq j,m\leq N\atop jm\leq N}\bigg[\frac{2p+m\beta^{-1}-j\beta }{2p-m\beta^{-1}+j\beta }\bigg]^{{\tt par}(N-mj)}\,,
\nonumber
\eea
where ${\tt par}(N)$ is the number of integer partitions of $N$.

\bigskip
Significant simplifications occur in the case when $\beta=\frac{1}{\sqrt{2}}$. Using the explicit formula
\eqref{Reig1b},\, \eqref{Reig1c} one has
\bea
R^{(1)}_p/R^{(0)}_p=\frac{P+1}{P-1}\ ,\ \ \ \ \ \ R^{(2,+)}_p/R^{(0)}_p=\frac{P+3}{P-1}\ ,\ \ \ \ 
R^{(2,-)}_p/R^{(0)}_p=\frac{P+1}{P-3}
\nonumber
\eea
 with $P=\sqrt{8}\, p$.  In general, for an arbitrary set ${\boldsymbol w}$,
 it is possible to show that
 \bea
 R({\boldsymbol w}_p)/R^{(0)}_p=\prod_{j=1}^J\frac{P+2n^{(-)}_j-1}{P-2n^{(+)}_j+1}\ .
 \nonumber
 \eea
Here $\{n_j^{(\pm)}\}$ are two sets of integers satisfying the condition
 $1\le n_1^{(\pm)}<n_2^{(\pm)}<\ldots <n_J^{(\pm)}$ and also
\be
N=\sum\limits_{j=1}^J\,\big(n^{(+)}_j+n^{(-)}_j-1\big)\ .
\nonumber
\ee
In fact, the sets $\{n_j^{(\pm)}\}$ can be used to classify  the  states $|{\boldsymbol w}\rangle$ 
for any $0<\beta\leq 1$. The integers $m_j$, which appear in the
exact Bohr-Sommerfeld quantization condition \eqref{jaasusaee}, are expressed through these numbers
(for details, see Appendix A in ref.\!\cite{Bazhanov:2003ni}).

\bigskip
Finally, it is worth noting that the spectral problem for
the reflection operator  \eqref{hssysys} turns out to be the most effective procedure for the explicit construction of 
the basis states  $|{\boldsymbol w}\rangle\in {\cal F}_p$.

\section{Inner product for the Verma module}

The formulae \eqref{sisoisss},\eqref{gsfss} combined with  the
Fourier expansion for $T(z)$ \eqref{Texpansion1}
imply that the local IM
can be understood as  elements of the universal enveloping algebra of $Vir$ without
any reference to the Heisenberg algebra.
Therefore  the diagonalization of the set $\{{\mathbb I}_{2n-1}\}_{n=1}^\infty$
can be formulated as a
problem in the Verma module rather than in the Fock space ${\cal F}_p$. 
Recall that  $|{\boldsymbol w}^{(N)}\rangle$  denotes  an eigenvector
of the local IM at level $N$ normalized by the condition \eqref{isisisis}. 
We now introduce the notation $\big|{\boldsymbol w}^{(N)}\big\rangle\!\!\big\rangle$  for the eigenvector
\bea\label{sisisisissu}
{\mathbb I}_{2n-1}\, \big|{\boldsymbol w}^{(N)}\big\rangle\!\!\big\rangle= I_{2n-1}\big({\boldsymbol w}^{(N)}\big)\, 
\big|{\boldsymbol w}^{(N)}\big\rangle\!\!\big\rangle\ ,
\eea
normalized such that
\bea\label{isisissis}
\big|{\boldsymbol w}^{(N)}\big\rangle\!\!\big\rangle=\big(\, (L_{-1})^N+\ldots \,\big)\, |v_p\rangle\ .
\eea
Here the dots denote the terms
involving the Virasoro algebra generators $L_{-n}$ with $2\le n\leq  N$.
The eigenvectors up to level $N=2$ are given by
\bea\label{sisisisnn}
\big|{\boldsymbol w}^{(0)}\big\rangle\!\!\big\rangle&=& |v_p\rangle\ ,\ \ \ \ \ \ \ \ \ 
\big|{\boldsymbol w}^{(1)}\big\rangle\!\!\big\rangle=L_{-1}\ |v_p\rangle\nonumber\\[-0.2cm]
\\[0cm]
\big|{\boldsymbol w}^{(2,\pm)}\big\rangle\!\!\big\rangle&=&
\Big( L_{-1}^2-\tfrac{1}{4}\, \big(2\rho^2+1\mp \sqrt{(2\rho^2-1)^2+32 p^2}\ \big)\ L_{-2}\,\Big)\,|v_p\rangle\ .\nonumber
\eea
Of course, being written in terms of the Heisenberg creation operators $\{a_{-n}\}_{n>0}$ via eq.\,\eqref{ososossi}, these states
coincide up to an overall factor with the highest Fock state $|p\rangle$,
$|{ \boldsymbol w}^{(1)}\rangle $
\eqref{sjssusu}
and $ \big|{\boldsymbol w}^{(2,\pm)}\big\rangle$ \eqref{aisaiisiass}, respectively.

\bigskip

The Virasoro algebra possesses the natural
Hermitian conjugation
\bea\label{sisisiss}
L_n^\star=L_{-n}
\eea
for any real values of the central charge $c=1+6Q^2$.
It is important to note that when 
 $Q$ is pure imaginary $Q=-\ri\rho$, i.e.,  $c< 1$,
this does not coincide with 
the Hermitian conjugation $a_n^\dagger=a_{-n}$ discussed 
in sec.\,\ref{sec6}.
The latter in this case, as it follows from eq.\,\eqref{ososossi}, does not lead
to a simple conjugation condition for $L_n$.
\bigskip

%Consider formulae \ \eqref{ososossi}, which express the Virasoro generators $L_n$ through
%the creation and annihilation operators $a_n$. When $Q$
%is pure imaginary, $Q=-\ri\rho$ , the Hermitian conjugation $a_n^\dagger=a_{-n}$ discussed 
%in Section \ref{sec6} does not lead to a simple conjugation condition for $L_n$.
%At the same time, the Virasoro algebra possesses the natural
%Hermitian conjugation
%\bea\label{sisisiss}
%L_n^\star=L_{-n}
%\eea
%for any real values of the center charge $c=1-6\rho^2$.

The formula \eqref{sisisiss} can be equivalently rewritten in terms of the holomorphic field $T(z)=T(x_2+\ri x_1)$:
\bea
T^\star(x_2+\ri x_1)=T(-x_2+\ri x_1)\ . \nonumber
\eea
In view of eqs.\eqref{sisoisss},\eqref{gsfss} this implies that
\bea\label{jasuusua}
{\mathbb I}_{2n-1}^\star={\mathbb I}_{2n-1}\ .
%\nonumber
\eea
Thus, the Hermitian conjugation \eqref{sisisiss} is consistent with the quantum KdV integrable structure.

\bigskip
The   Verma module  has a unique Hermitian form  that is
 induced from the conjugation \eqref {sisisiss} and such that
the norm of the highest vector is one. Because of the Hermiticity of the
local IM  \eqref{jasuusua}, this Hermitian form is diagonal
in the basis of the eigenstates $\big|{\boldsymbol w}\big\rangle\!\!\big\rangle$ \eqref{sisisisissu},\,\eqref{isisissis}:
\bea\label{isiiisaisaias}
\big\langle\!\!\big\langle {\boldsymbol w}\big|{\boldsymbol w}'\big\rangle\!\!\big\rangle_{Vir}=V_p({\boldsymbol w})\ \delta_{{\boldsymbol w},
{\boldsymbol w}'}\ .
\eea

\bigskip

For  example, for the states \eqref{sisisisnn} at the first  few levels one has
\bea\label{oiwoioiuds1}
&&V_p\big({\boldsymbol w}^{(0)}\big)=1\, ,\ \ \ \ 
V_p\big({\boldsymbol w}^{(1)}\big)=\half\ (2p-\rho)(2p+\rho)\nonumber \\[-0.2cm]
\\[0.1cm]
%G\big(\bm{w}^{(2,\pm)}\big)&=&16p^4+(\tfrac{5}{2}-18\rho^2+2\rho^4)\,p^2+
%\tfrac{1}{16}\,(1-2\rho^2)^2\,(1-8\,\rho^2) \\[0.4cm]
%&\mp&\tfrac{1}{16}\, \big(1+6\rho^2-16\rho^4+8p^2\,(2\rho^2-5)\big)\, \sqrt{(2\rho^2-1)^2+32 p^2}\  .\nonumber
&&V_p\big(\bm{w}^{(2,\pm)}\big)=\mp\, \tfrac{1}{64}\,B\,\big(1+2\rho^2\mp B\big)\,
\big(3+2\beta^2-2\beta^{-2}\pm B\big)\,\big(3-2\beta^2+2\beta^{-2}\pm B\big)\ ,\nonumber
\eea
where $B=\sqrt{(2\rho^2-1)^2+32 p^2}$.
Notice that the product of $V_p(\bm{w}^{(N)})$ 
for all the states with given $N$ coincides with
the Gram determinant for \eqref{isiiisaisaias}  restricted to this level.
The calculation for the lowest levels leads to
%Thus, the famous Kac formula \cite{} leads to
\bea
\prod_{{\boldsymbol w^{(N)}}}V_p\big({\boldsymbol w}^{(N)}\big)=D^{(N)}(p)\ 
\prod_{1\leq j, m\leq N\atop jm\leq N}\big((2 p+m\beta^{-1}-j\beta)
(2 p-m\beta^{-1}+j\beta)\big)^{{\tt par}(N-jm)}\ . \nonumber
\eea
This is reminiscent of the Kac determinant  formula \cite{Kac}. However, contrary to the latter, the factor $D^{(N)}(p)$ does  depend on $p$
and $\rho$  for $N\geq 2$.
In particular
\bea
D^{(0)}=1,\ \ \ \ D^{(1)}=\half,\ \ \ \ D^{(2)}=\tfrac{1}{8}\, B^2=\tfrac{1}{8}\, \big((2\rho^2-1)^2+32p^2\big)\ .\nonumber
\eea
%$A_N$ is a positive constant that does not depend on $p$ and $\beta$.

\bigskip
In the l.h.s. of  eq.\eqref{isiiisaisaias} we use the subscript $Vir$ to emphasize that the inner product is calculated
using the natural Hermitian conjugation \eqref{sisisiss} for the Virasoro algebra. On the other hand, one can treat
  $\big|{\boldsymbol w}\big\rangle\!\!\big\rangle$ and $\big|{\boldsymbol w}'\big\rangle\!\!\big\rangle$  as
  states from  the Fock space ${\cal F}_p$ and calculate their inner product using the Hermitian conjugation
 $a_n^\dagger=a_{-n}$ for the Heisenberg algebra. The result, of course, is different:
 \bea\label{isiiisaisaias1a}
\big\langle\!\!\big\langle {\boldsymbol w}\big|{\boldsymbol w}'\big\rangle\!\!\big\rangle_{Heis}=H_p({\boldsymbol w})\ \delta_{{\boldsymbol w},
{\boldsymbol w}'}\ .
\eea
 The relations \eqref{ossisisi},\,\eqref{sosoisis} imply that
 \bea\label{ssososos}
 H_p({\boldsymbol w})=r_p({\boldsymbol w})\, V_p({\boldsymbol w})\ \ \ \  \ {\rm with}\ \ \ \ \ r_p({\boldsymbol w})=
 R_p({\boldsymbol w})/R^{(0)}_p\ .
 \eea

 \bigskip
 \section{\label{sec10}Star conjugation for the finite spin chain}

 We now turn to the main subject of this paper -- the norm ${\mathfrak N}$  \eqref{aiusauasssu}, or 
 equivalently  \eqref{aiusssu}. In fact, instead of ${\mathfrak N}$ 
 it is useful to focus on the  combination
 \bea\label{osissisissd}
 {\mathfrak  K}={ \mathsf U}^{-1}\  {\mathfrak N}\ .
 \eea
 Here
 %\bea
 %J=\prod_{j=1}^M 4\sinh\big(\beta_j+\tfrac{\ri\gamma}{2}\big)\sinh\big(\beta_j-\tfrac{\ri\gamma}{2}\big)
 %\eea
 \bea\label{ososaoosa}
 { \mathsf U}=(-1)^M\ \big(q-q^{-1}\big)^{-2M}\  \prod_{m=1}^M \big(1+\zeta_m\, q\big)\big(\zeta_m^{-1}+q^{-1}\big)
 \eea
 and relations \eqref{prodeqA},\,\eqref{prodeqB}
imply the following scaling behavior for this quantity
 \bea\label{isaiisais}
 { \mathsf U }=R_{ \bar p}({\bar {\boldsymbol w}})\,R_{ p}({\boldsymbol w}) \ 
 \big(2\sin(\pi\beta^2)\big)^{\frac{2(p+{\bar p})}{\beta}}
\bigg(\frac{L}{\pi}\bigg)^{-2\rho\, ({\bar p}+{p})}\ 
\bigg(\frac{2(1-\beta^2)}{\sin(\pi\beta^2)}\bigg)^{L}\,\big(1+o(1)\big)\ .
 \eea
 Combining these with eqs.\eqref{aososaik},\eqref{asdasd1122},\eqref{ossiss},\eqref{scalingN1}
 one finds
 \bea\label{aoso1a}
{ \mathfrak K}={\cal K}_{\bar p}({\bar {\boldsymbol w}})\, {\cal K}_p({\boldsymbol w})\  L^{\eta_K}\ \re^{{\cal A}_2 L^2}\ \big(1+o(1)\big)
\eea
with the scaling exponent given by
  \bea\label{jsyssts}
 \eta_K=\tfrac{1}{6}-\tfrac{1}{3}\,\rho^2
 -4 \,I_1({\boldsymbol w})-4 \,{\bar I}_1({\bar {\boldsymbol w}})\ .
 \eea
The amplitudes ${\cal K}_{\bar p}({\bar {\boldsymbol w}})$ and ${\cal K}_p({\boldsymbol w})$
 are related to ${\cal N}_{\bar p}({\bar {\boldsymbol w}})$ and ${\cal N}_p({\boldsymbol w})$ in
 \eqref{scalingN1} as
 \bea\label{qqssisiis}
 {\cal N}_{\bar p}({\bar {\boldsymbol w}})&=& 
 \big(2\sin(\pi\beta^2)\big)^{\frac{2{\bar p}}{\beta}}\  \pi^{2\rho {\bar p}}\ R_{ \bar p}({\bar {\boldsymbol w}})\
 {\cal K}_{\bar p}({\bar {\boldsymbol w}})\nonumber\\[-0.2cm]
&&\phantom{s}\hskip6cm\ .
 \\[-0.2cm]
 {\cal N}_p({\boldsymbol w})&=&\big(2\sin(\pi\beta^2)\big)^{\frac{2{ p}}{\beta}}\ \pi^{2\rho { p}}\ 
 R_{ p}({\boldsymbol w})\  {\cal K}_p({\boldsymbol w})\nonumber
 \eea

 \bigskip
 The motivation for studying ${\mathfrak K}$ comes from 
Korepin's approach to the norms in ref.\!\cite{Korepin:1982ej}, which is based on the
Quantum Inverse Scattering Method. 
In this framework,
the main player is
the quantum monodromy matrix
that is built from the ordered product of the elementary transport matrices:
\bea\label{soosss}
%{\hat {\boldsymbol M}}(\theta)\equiv
\begin{pmatrix}
{\hat {\mathsf A}}(\theta)&{\hat {\mathsf B}}(\theta)\\
 {\hat {\mathsf C}}        (\theta)&{\hat {\mathsf D}}(\theta)
         \end{pmatrix}={\overset{\leftarrow}{\cal P}}\, \prod_{n=1}^{L}
 \begin{pmatrix}
\sinh(\theta-\frac{\ri\gamma}{2}\,\sigma_n^z)&-\ri \sin(\gamma)\, \sigma^-_n\\
-\ri\sin(\gamma)\, \sigma^+_n&\sinh(\theta+\frac{\ri\gamma}{2}\,\sigma_n^z)
\end{pmatrix} \ .
\eea
In particular, provided the following identifications are made
\bea\label{param1a}
\zeta=\re^{-2\theta}\ ,\ \ \ \ \ \ \ \ \zeta_m=\re^{-2\theta_m}\ ,\  \ \ \ \ \ q=-\re^{-\ri\gamma}\ ,
\eea
and
taking into the account that the Hamiltonian
in ref.\!\cite{Korepin:1982ej} differs from  \eqref{asiisaias}
by  means of the similarity transformation with the matrix\footnote{In this formula $L$ is assumed to be even.}
\bea
\hat{\mathsf{\Sigma}}=\prod_{n=1}^{\frac{L}{2}} \sigma^z_{2n-1} \ ,
\nonumber
\eea
the Bethe state corresponding to the wave function \eqref{aisaiasu},
\bea
|\Psi\rangle=
\sum_{1\leq  x_1<x_2<\ldots<x_M\leq L}\Psi(x_1,\ldots,x_M)\, \sigma_{x_1}^-\cdots
 \sigma_{x_M}^-\,|\,0\,\rangle\ ,\nonumber
\eea
can be  nicely expressed  in terms of the operators\footnote{In the l.h.s. of this relation   we use the argument $\zeta$ instead of $\theta$.
It is easy to see that the operator
${\hat {\boldsymbol {  \mathsf B}}}$ is a   single valued function of
this variable.}
\bea\label{lsososos}
{\hat {\boldsymbol {  \mathsf B}}}(\zeta)\equiv 
\ri\ 
\frac{\sinh\big(\theta+\tfrac{\ri\gamma}{2}\big)}{\sin(\gamma)}\ 
\Big[\sinh\big(\theta-\tfrac{\ri\gamma}{2}\big)\Big]^{-L}\
\hat{\mathsf{\Sigma}}\,  {\hat {\mathsf B}}(\theta)\hat{\mathsf{\Sigma}}\ .
\eea
Namely, one has 
\bea\label{owieoi12}
|\Psi\rangle=%
%\big(\!-4\ri\sin(\gamma)\big)^{-M}\ \ U\ 
\prod_{m=1}^M 
{\hat {\boldsymbol { \mathsf  B}}}(\zeta_m)\,|\,0\,\rangle\ .
\eea
Here $|0\rangle$ stands for the pseudo-vacuum 
\bea
|\,0\,\rangle=\underbrace{
\begin{pmatrix}1\\0\end{pmatrix}
\otimes\cdots \otimes\begin{pmatrix}1\\0
\end{pmatrix}
}_{L\ {\rm times}}\ .
\nonumber
\eea
%and $U$ is the same as in \eqref{ososaoosa}. 
%Within the parameterization \eqref{param1a} it reads as
%\bea
%U=\prod_{m=1}^M 4\sinh\big(\theta_m+\tfrac{\ri\gamma}{2}\big) 
%\sinh\big(\theta_m-\tfrac{\ri\gamma}{2}\big)\ .
%\nonumber
%\eea
Notice that the states \eqref{owieoi12} diagonalize  the transfer-matrix
\bea\label{isisisiss}
{\hat {\boldsymbol {  \mathsf T}}}(\zeta)=\re^{-\theta L}\ 
\hat{\mathsf{\Sigma}}\,
\big(\re^{\ri\pi{\tt k}}\  { \mathsf  A}(\theta)+\re^{-\ri\pi{\tt k}}\  { \mathsf  D}(\theta)\big)\hat{\mathsf{\Sigma}}\ .
\eea
The latter  is related to the Hamiltonian \eqref{asiisaias} as
\bea
 {  \mathsf H}\,
=
2\, \big(q-q^{-1}\big)\ \zeta\partial_\zeta\log {\hat {\boldsymbol {  \mathsf T}}}(\zeta)\big|_{\zeta=-q^{-1}}+
\half\ (3q^{-1}-q)\,L\ .\nonumber
\eea

\bigskip
Using eq.\,\eqref{owieoi12}  and that 
$\{\zeta_j\}_{j=1}^M$ coincides with the complex conjugated set,
the norm ${\mathfrak N}\equiv\langle \Psi|\Psi\rangle$ can be
represented as
\bea\label{sososos}
{\mathfrak N}=
%\big(4\sin(\gamma)\big)^{-2M}\ U^2\ 
\big\langle\,0\,\big| \prod_{j=1}^M 
\big( {\hat {\boldsymbol { \mathsf B}}}(\zeta^*_j)\big)^\dagger\prod_{m=1}^M 
 {\hat {\boldsymbol {  \mathsf B}}}(\zeta_m)\,\big|\,0\,\big\rangle\ .
\eea
On the other hand, a simple calculation
based on  the definitions \eqref{soosss} and \eqref{lsososos} shows that
\bea
\big({\hat {\boldsymbol {  \mathsf B}}}(\zeta^*)\big)^\dagger={\hat {\boldsymbol {  \mathsf C}}}(\zeta)\ ,
\nonumber
\eea
where the operator  ${\hat {\boldsymbol {\mathsf C}}}(\zeta)$ is defined   similarly to 
 \eqref{lsososos}:
\bea\label{lsososos1a}
{\hat {\boldsymbol {  \mathsf C}}}(\zeta)\equiv
\ri\ 
\frac{\sinh\big(\theta-\tfrac{\ri\gamma}{2}\big)}{\sin(\gamma)}\ 
\Big[\sinh\big(\theta+\tfrac{\ri\gamma}{2}\big)\Big]^{-L}\ \
  \hat{\mathsf{\Sigma}}\,{\hat {\mathsf C}}(\theta)\hat{\mathsf{\Sigma}}\ .
  \nonumber
\eea
Thus eq.\eqref{sososos} takes  the form
\bea\label{Isososos}
{\mathfrak N}=
%\big(4\sin(\gamma)\big)^{-2M}\ U^2\ 
%(-1)^M\ 
\big\langle\,0\,\big| \prod_{j=1}^M 
{\hat {\boldsymbol { \mathsf C}}}(\zeta_j)\prod_{m=1}^M 
 {\hat {\boldsymbol {  \mathsf B}}}(\zeta_m)\,\big|\,0\,\big\rangle\ .
\eea
In ref.\!\cite{Gaudin:1981cyg} Gaudin, McCoy and Wu  conjectured that
the above norm coincides with the r.h.s. of eq.\,\eqref{aiusauasssu} with $f$
taken as in eq.\eqref{kaajay}.
This was proven in the work \cite{Korepin:1982ej}.

\bigskip
Let us rewrite eq.\eqref{Isososos}
in terms of  ${ \mathfrak   K}$  defined by \eqref{osissisissd}:
\bea\label{keqoi132}
{\mathfrak K}=
%(-1)^M\ \big(q-q^{-1}\big)^{2M}\ { U}^{-1}\ 
{ \mathsf  U}^{-1}\ 
\big\langle\,0\,\big| \prod_{j=1}^M 
{\hat {\boldsymbol { \mathsf C}}}(\zeta_j)\prod_{m=1}^M 
 {\hat {\boldsymbol {  \mathsf B}}}(\zeta_m)\,\big|\,0\,\big\rangle\ .
\eea
We now  note  that  ${ \mathsf U}$  \eqref{ososaoosa} coincides with the eigenvalue of
the operator
\bea\label{ssssyys}
{\hat {\mathsf U}}= \big(\ri\, (q-q^{-1})\big)^{2{\hat  {\mathsf S}^z-L}}\ 
{\hat {\boldsymbol {\mathsf Q}}}(-q)
 \,{\hat {\boldsymbol {\mathsf Q}}}(-q^{-1})\ \big[{\hat {\boldsymbol {\mathsf Q}}}(0)\big]^{-1}
\eea
 built from the $Q$-operator.
 Then with  eq.\,\eqref{keqoi132} one can interpret ${ \mathfrak   K}$ to be the norm of the Bethe state
 $\prod_{m=1}^M 
 {\hat {\boldsymbol {  \mathsf B}}}(\zeta_m)\,\big|\,0\,\big\rangle$
 w.r.t.  the ``star'' conjugation,
 which is 
 related to the ordinary matrix conjugation ${\hat {\mathsf O}}^\dagger
\equiv \big({\hat { \mathsf O}}^T\big)^*$
as 
\bea\label{iiisisaas}
{\hat {\mathsf O}}^\star={\hat{\mathsf U}}\, {\hat { \mathsf O}}^\dagger\, {\hat{\mathsf U}}^{-1}\ .
\eea
In  view of eqs.\eqref{isiiisaisaias},\,\eqref{isiiisaisaias1a},\,\eqref{ssososos},\,\eqref{osissisissd},\,\eqref{isaiisais}
the star conjugation \eqref{iiisisaas} for the spin chain with 
 finite length becomes    ${\bar L}_n^\star={\bar L}_{-n}$ and $L_n^\star=L_{-n}$ 
 for $\overline{Vir}\otimes
 { Vir}$ in the scaling limit.

\bigskip

For any operator commuting with ${\hat {\boldsymbol {\mathsf Q}}}(\zeta)$, one has that
$\hat{{\mathsf O}}^\star=\hat{{\mathsf O}}^\dagger$.
In particular, it is easy to show that for the transfer-matrix \eqref{isisisiss} 
 \bea
 \big({\hat {\boldsymbol {  \mathsf T}}}(\zeta)\big)^\star=\big({\hat {\boldsymbol {  \mathsf T}}}(\zeta)\big)^\dagger=
 {\hat {\boldsymbol {  \mathsf T}}}(\zeta^*)\ .
 \eea
 When the twist parameter ${\tt k}\not=0$, the transfer-matrix is expected to lift all the degeneracies
 in  the basis of  the stationary states, so  that $\prod_{m=1}^M 
 {\hat {\boldsymbol {  \mathsf B}}}(\zeta_m)\,\big|\,0\,\big\rangle$ corresponding to different sets $\{\zeta_m\}_{m=1}^M$
 turn out to be orthogonal w.r.t. the inner products associated with both the ``dagger'' and ``star'' conjugations. 
Finally, we note that
the norm corresponding to the star conjugation is not positive definite.

\section{Outcomes of numerical work}
Unfortunately we don't know of any
analytic techniques that could allow one to derive
the scaling behavior  of the norms of the Bethe states
\eqref{aososaik}, including explicit expressions 
for the scaling exponent $\eta$ and 
amplitude ${\mathfrak N}_\infty$. 
However, through numerical work, we
conjectured the formula \eqref{ossiss} for $\eta$,
and found that  ${\mathfrak N}_\infty$ obeys the factorized structure \eqref{scalingN1}. The latter, as
a consequence of eq.\eqref{qqssisiis} 
can be written in the form
\bea\label{scalissfs}
{\mathfrak N}_\infty =\big(2\sin(\pi\beta^2)\big)^{2 S^z}\ 
 \big[\,\pi^{2\rho{\bar p}}\, R_{ \bar p}({\bar {\boldsymbol w}})\,{\cal K}_{\bar p}({\bar {\boldsymbol w}})\big]\ 
\big[\,\pi^{2\rho{ p}}\,R_{ p}({\boldsymbol w}) \,  {\cal K}_p({\boldsymbol w})\big]\ .
 \eea
Recall that $2p=\beta\,S^z+\beta^{-1} \,({\tt k}+{\tt w})$, 
$2\bar{p}=\beta\,S^z -\beta^{-1}\, ({\tt k}+{\tt w})$ and $\rho=\beta^{-1}-\beta$.
In sec.\,\ref{sec8}, we discussed how to compute  the eigenvalues $R_{ p}({\boldsymbol w})$ ($R_{ \bar p}({\bar {\boldsymbol w}})$)
 of the
reflection operators $\mathbb{R}$ $(\bar{\mathbb{R}})$
entering into the above formula. 
Thus, we are left to describe  the unknown factors ${\cal K}_p({\boldsymbol w})$ and ${\cal K}_{\bar p}({\bar {\boldsymbol w}})$.
Since there is only a nomenclature difference
between the two,  it is sufficient to focus on ${\cal K}_p({\boldsymbol w})$ only.
\bigskip

In order to present the results of our numerical study, we will first define 
some special functions  and briefly describe their properties.

\subsection{The special functions $Z_\pm(p\,|\,\beta)$}

Introduce the notation
\bea
h(p)=4 p^2+\tfrac{1}{6}\, (\rho^2-1)\,.
\eea
Notice that the scaling exponent $\eta_K$ \eqref{jsyssts} is given by
\bea
\eta_K=\tfrac{1}{6}-h(p)-h({\bar p})\ -4\,(N+\bar{N})\ .
\nonumber
\eea
Then, the function $Z_+$ is defined through the convergent product
\bea\label{asusausss}
Z_+(p\,|\,\beta)&=&(A_{\rm G})^{-2 \beta^2}\ 
(2\pi)^{\frac{1}{2}-2p\beta}\ \beta^{h(p)+4p\beta+1}\ 
\frac{
\re^{-(\frac{2p}{\beta}+h({ p})+\frac{1}{2}-\frac{1}{6}\, \beta^2)\gamma_{\rm E}}
}{\Gamma(1+\frac{2p}{\beta})\, \Gamma(1+2p\beta)}\nonumber\\[0.2cm]
&\times&
 \prod_{m=1}^\infty
\frac{2\pi\, (m\beta^2)^{2m\beta^2+4p\beta+1}\, \re^{-2m\beta^2+\frac{1}{m}(\frac{2p}{\beta}
+h({ p)}+\frac{1}{2}-\frac{1}{ 6}\beta^2)}}{\Gamma^2(1+2p\beta+m \beta^2)}\ .
\eea
%\bea\label{asusau}
%\small{G_+({\tt p}|g)
%&=&(2\pi)^{\frac{g}{2}-{\tt p}}\  A^{-2g}_G\ 
%\re^{\gamma_ E\big({\tt p}-\omega({\tt p})-\frac{g}{3}
%\big)}
%\
%\prod_{m=1}^\infty
%\frac{2\pi\,  (g m)^{2{\tt p}+g(2m-1)}\  \re^{-2m g-\frac{{\tt p}}{m}+\frac{\omega({\tt p})}{gm}+\frac{g}{3m}
%}
%}{\Gamma(1+{\tt p}+(m-1) g)\,\Gamma\big({\tt p}+g m\big)}\\[0.2cm]
%&=&
%=A_G^{-2 g}\, \bigg(\frac{g}{2\pi}\bigg)^{\tt p}\ \frac{\sqrt{2\pi g}
%\, \re^{-\gamma_E(\frac{p}{g}+\omega({\tt p})+\frac{1}{2}-\frac{g}{6})}
%}{\Gamma(1+\frac{\tt p}{g})\, \Gamma(1+{\tt p})}\, 
 %\prod_{m=1}^\infty
%\frac{2\pi\, (mg)^{2mg+2{\tt p}+1}\, \re^{-2mg+\frac{1}{m}(\frac{p}{g}
%+\omega({\tt p)}+\frac{1}{2}-\frac{g}{ 6})}}{\Gamma^2(1+{\tt p}+m g)}}
%\eea
Here $A_{\rm G}$ and $\gamma_{\rm E}$ are  the Glaisher and Euler constants, respectively, while
 $\beta$ plays the r${\hat {\rm o}}$le of the parameter which is always assumed to be positive. 
 %To simplify the notation, we will often drop the dependence on  $\beta$:
% \bea
%G_+(p)\equiv G_+(p\,|\,\beta) \ .
%\eea 
It follows from the definition that  $Z_+({ p}\,|\,\beta)$  is an entire
function of ${ p}$ whose 
zeros  are located on the real  negative semi-axis at the points
\bea
2{ p}=
\begin{cases}
-m\beta^{-1}\,, -m\beta\ \ \ \ \ \ \ &{\rm for}\ m=1,2,\ldots\ \ \ \ \  \ \ \ \ ({\rm simple\  zero})\\[0.2cm]
-n\beta^{-1}-m \beta\ \ \ \ \ \ \ &{\rm for}\ m,n=1,2,\ldots\ \ \ \ \  \ ({\rm double\   zero})
\end{cases}\ .
\nonumber
\eea
As ${p}\to +\infty$ the function $Z_+({p}\,|\,\beta)$ possesses  the  asymptotic behavior
\bea
\log Z_+({ p}\,|\,\beta)=h({ p})\ \log\big({2p} \, \re^{-\frac{3}{2}}\,\big)+O(1)\ .
\nonumber
\eea
In many situations it is convenient to use  the integral representation
\bea\label{iaasssa}
Z_+({p}\,|\,\beta)=\exp\Bigg[
-\int_0^\infty\frac{\rd t}{t}\bigg(\, \frac{\re^{-4p t}\,\cosh(\rho t)}{2\sinh(\beta t)
\sinh(\frac{t}{\beta})}-\frac{1}{2t^2}+\frac{2p }{t}
-h({ p})\, \re^{-2 t} \bigg)\Bigg]\,,
\eea
which is valid in the half plane $\Re  e({2p})>-\min(\beta,\beta^{-1})$.
Note that this formula immediately implies that $Z_+({ p}\,|\,\beta)$   satisfies  the ``duality'' relation
\bea
Z_+\big( p\,\big|\beta^{-1}\big)=Z_+(p\,|\,\beta)\ .
\nonumber
\eea

Similar to \eqref{iaasssa}, we define another function
$Z_-({ p}\,|\,\beta)$:
\bea\label{jsssss}
 Z_-({p}\,|\,\beta)=\exp\Bigg[
\int_0^\infty\frac{\rd t}{t}\
\Bigg(\, \frac{\re^{-4pt}\,\sinh(\rho t)}{2\sinh(\beta t)\sinh(\frac{t}{\beta})}-\frac{\rho}{2t}
+2p\rho \   \re^{-2t}\, \Bigg)\Bigg]\ ,
\eea
that obeys the relation
\bea
Z_-\big( p\,\big|\beta^{-1}\big)=\big[Z_-(p\,|\,\beta)\big]^{-1}\ .
\nonumber
\eea
Contrary to  \eqref{iaasssa} it is possible to explicitly evaluate the integral in eq.\eqref{jsssss}, which yields
\bea
Z_-({p}\,|\,\beta )=\beta^{1+2p(\beta^{-1}+\beta)} \ \frac{\Gamma(1+\frac{2p}{\beta})}{\Gamma(1+2p\beta)}\ .\nonumber
\eea
Notice that,
using the function $Z_-({p}\,|\,\beta )$, eq.\eqref{Reig1a} can be rewritten as
\bea\label{osossis}
 \pi^{2p\rho}\ R^{(0)}_p= \,
 \Bigg[\frac{\sqrt{\pi} \Gamma(1+\frac{\beta}{2\rho})}{\beta\, \Gamma(\frac{3}{2}+\frac{\beta}{2\rho})}
\Bigg]^{2p\rho}\ \, Z_-({p}\,|\,\beta )\ 
 \ .
 \eea
 
 \bigskip
 Finally we define the entire function 
 \bea\label{sususus}
 Z({ p}\,|\,\beta)=\sqrt{Z_+({ p}\,|\,\beta)Z_-({ p}\,|\,\beta)}\ ,
 \eea
which for  $\Re  e({2p})>-\beta^{-1}$ admits the integral representation
\bea\label{iaasssa11}
\!\!\!\!\!\!\! Z({p}\,|\,\beta)=\exp\Bigg[\!
-\!\int_0^\infty\! \frac{\rd t}{2t}\bigg(\, \frac{\re^{-(4p+\rho) t}}{2\sinh(\beta t)
\sinh(\frac{t}{\beta})}-\frac{1}{2t^2}+\frac{(4p+\rho) }{2t}
-\big(h({ p})+2p\rho\big)\, \re^{-2 t} \bigg)\!\Bigg]\, .\nonumber
\eea
For any complex $p$ the value of  $Z({ p}\,|\,\beta)$ can be found by means of  the convergent product
\bea\label{asss}
Z(p\,|\,\beta)&=&(A_{\rm G})^{- \beta^2}\ 
(2\pi)^{\frac{1}{4}-p\beta}\ \beta^{\frac{1}{2} h(p)+1+p(\beta^{-1}+3\beta)}\ 
\frac{
\re^{-\frac{1}{2} (\frac{2p}{\beta}+h({ p})+\frac{1}{2}-\frac{1}{6}\, \beta^2)\gamma_{\rm E}}
}{\Gamma(1+2p\beta)}\nonumber\\[0.2cm]
&\times&
 \prod_{m=1}^\infty
\frac{\sqrt{2\pi}\, (m\beta^2)^{m\beta^2+2p\beta+\frac{1}{2}}\, \re^{-m\beta^2+\frac{1}{2m}(\frac{2p}{\beta}
+h({ p)}+\frac{1}{2}-\frac{1}{ 6}\beta^2)}}{\Gamma(1+2p\beta+m \beta^2)}\ .\nonumber
\eea
\newpage

\subsection{Conjectures}

%In  the case of the primary Bethe states, which correspond to the level  $N=\bar{N}=0$,
% ${\boldsymbol w}^{(0)}$ is the empty set. It is convenient to use the shortcut notation
%${\cal K}^{(0)}_p\equiv {\cal K}_p({\boldsymbol w}^{(0)})$.
Through a numerical study of the norms in the scaling limit, 
we were lead to the following
\bigskip
%\medskip

 {\bf Conjecture II:} 
In  the case of the primary Bethe states, which correspond to the level  $N=\bar{N}=0$,
\bea\label{jsasauysa1a}
{\cal K}_p({\boldsymbol w}^{(0)})=C_0\ C^{h(p)}\ Z_+(p\,|\,\beta)
%\big(\re^{-1}{\mathfrak C}\big)^{-\frac{{\tt p}^2}{g}}=
%\bigg(\frac{C_+}{\beta}\bigg)^{4p^2}\ \frac{Z_+(p\,|\,\beta)}{Z_+(0\,|\,\beta)}
\ .
\eea
Here the constants ${ C}_0, \, C$  are independent of ${p}$ and depend only on
the parameter $\beta$.

\begin{figure}
\centering
\scalebox{1.0}{
\begin{tikzpicture}
%\node at (-3.3,2.8) {\small$\log{\mathfrak A}(g)$};
\node at (-3.4,2.7) {\small$\log(\beta C)$};
\node at (4.3,-2.0) {\small $\beta^2$};
%\node at (14.3,-2.0) {\small $g$};
\node at (0,0) {\includegraphics[width=0.45\textwidth]{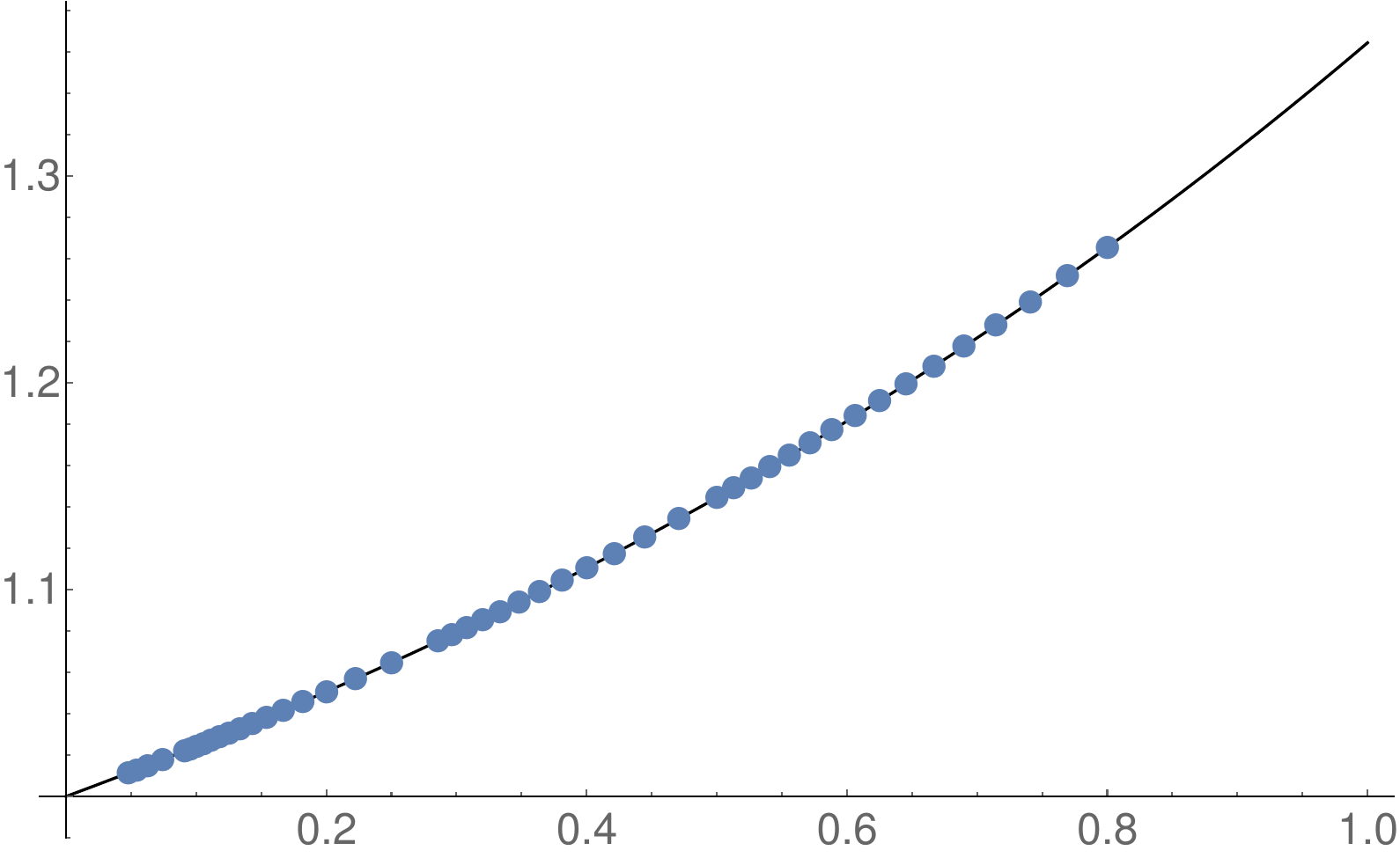}};
%\node at (10,0) {\includegraphics[width=0.45\textwidth]{CB}};
\end{tikzpicture}
}
\caption{\small\label{fig6}The  logarithm of  the $p$-independent constant 
$\beta C$  as a function of the parameter $\beta^2$.  
The dots correspond to the numerical data extracted from eqs.\,\eqref{jsasauysa1a}\,\eqref{aoso1a}, where
the l.h.s. was computed from the solution of the BA equations.  The solid line represents the cubic
 fit 
$\log (\beta C)^{(\rm fit)}= 1+
0.2334\, \beta^2 +0.091\, \beta^4 + 0.04\, \beta^6$.
Notice that 
  $\log(\beta  C)|_{\beta=0}=1$, while
$\log (\beta C)|_{\beta=\frac{1}{\sqrt{2}}}=\log(\pi)=1.14473$. Interpolation of the data yields that 
$\log(\beta C)|_{\beta=1}= 1.36{\bar 4}$ .}
%        $ (  \log(2)=0.693147)$. }
\end{figure}

\bigskip
Fig.\,\ref{fig6}   depicts $\log (\beta C)$  as a function of $\beta^2$.
Notice that
for  $\Delta=0$ the  $XXZ$ spin $\frac{1}{2}$  chain can be reformulated 
as a non-interacting  system of  1D lattice fermions.
In this case, $\beta=\frac{1}{\sqrt 2}$
and one can show that
\bea
\beta C\big|_{\beta=\frac{1}{\sqrt{2}}}=\pi\ .
\nonumber
\eea
%Moreover, our  numerical work suggests that
%\bea
%\lim_{\beta\to  0}\beta C=\re\ . 
%\nonumber
%\eea
We found it useful to re-write
 the other constant  $C_0$ appearing in eq.\,\eqref{jsasauysa1a} in the form
\bea
C_0=\beta^{-\frac{1}{3}}\  \re^{-\frac{1}{6}\rho^2 }\  
\bigg(\frac{\re^{\gamma_{\rm E}+1}}{4\pi}\bigg)^{\frac{1}{12}}\ 
\big(\re^{-\frac{1}{6}(\gamma_{\rm E}+1)}\ A^2_{\rm G}\big)^{\beta^2}\ {\tilde C}_0\ .
\eea
Then for the free fermion case
\bea
{\tilde C}_0|_{\beta=\frac{1}{\sqrt{2}}}=1\ .
\nonumber
\eea
Moreover, within the accuracy of our calculations,  ${\tilde C}_0\approx1$ in the domain $0.06<\beta^2<0.8$.

\bigskip
\newpage

 {\bf Conjecture III:} 
For a general Bethe state
\bea\label{jsasauysa}
{\cal K}_p({\boldsymbol w})=C_0\ C^{h(p)+4N}\ V_p({\boldsymbol w})\, Z_+(p\,|\,\beta)
%\big(\re^{-1}{\mathfrak C}\big)^{-\frac{{\tt p}^2}{g}}=
%\bigg(\frac{C_+}{\beta}\bigg)^{4p^2}\ \frac{Z_+(p\,|\,\beta)}{Z_+(0\,|\,\beta)}
\ ,
\eea 
where $V_p({\boldsymbol w})$ is defined by eqs.\,\eqref{isiiisaisaias},\,\eqref{sisisisissu},\,\eqref{isisissis}
 and the constants
$C_0,\,C$ are the same as in eq.\,\eqref{jsasauysa1a}.
This formula was numerically verified for
$N=0,1,2$ (for an illustration, see tabs.\,\ref{tab1} and \ref{tab2}).
% For all these cases we found that $C_N$ coincided with $C_0$.
%In general, for $N\ge 3$, $C_N$ is expected to be
%some $\beta$-independent constant and we
%do not exclude the possibility that it may depend on the level.

\bigskip

Eq.\eqref{jsasauysa} supplemented  by  \eqref{scalissfs},\,\eqref{ssososos},\,\eqref{osossis},\,\eqref{sususus}
implies the following result for the scaling amplitude ${\mathfrak N}_\infty$ from \eqref{aososaik}:
\bea\label{isussuus}
{\mathfrak N}_\infty&=& C^2_0\ C^{h(p)+h({\bar p})+4(N+\bar{N})}  \ \big(2\sin(\pi\beta^2)\big)^{\frac{2}{\beta} (p+{\bar p})}\ 
 \Bigg[\frac{\sqrt{\pi} \Gamma(1+\frac{\beta}{2\rho})}{\beta\, \Gamma(\frac{3}{2}+\frac{\beta}{2\rho})}
\Bigg]^{2(p+{\bar p}) \rho } \nonumber \\[0.3cm]
&\times &H_p(\bm{w})\,H_{\bar p}(\bar{\bm{w}})\,\ Z^2(p)\, Z^2({\bar p})\ .
\eea
The state dependent factor  $H_p(\bm{w})$ is unambiguously defined by eqs.\,\eqref{sisisisissu},\,\eqref{isisissis},\,\eqref{isiiisaisaias1a}.
For the primary Bethe states $(N=\bar{N}=0)$ one has that $H_p(\bm{w}^{(0)})=1$.

\begin{table}
\centering
\scalebox{0.85}{
\begin{tabular}{|c|c|c|c|c|c|}
\hline
 & & & & & \\[-0.3cm]
 Parameters & $L=50$ & $L=100 $ & $L=200$ & $L=\infty$ & $V_p(\bm{w}^{(1)})$ \\
\hline
 & & & & & \\[-0.3cm]
$\beta^2=\frac{2}{5},\, S^z={\tt w}=0,\, \pi{\tt k}=0.55$ & $-0.41042644$ & $-0.41137298$ & $-0.41160999$ & $-0.41168899$ & $-0.41168793$ \\[0.05cm]
\hline
 & & & & & \\[-0.3cm]
$\beta^2=\frac{10}{23},\, S^z={\tt w}=0,\, \pi{\tt k}=-0.12$ & $-0.36457515$ & $-0.36542880$ & $-0.36564253$ & $-0.36571378$ & $-0.36571343$ \\[0.05cm]
\hline
 & & & & & \\[-0.3cm]
$\beta^2=\frac{4}{9},\, S^z=1,\,{\tt w}=0,\, \pi{\tt k}=0.4$  & 
$\phantom{-}0.02054114$ & $\phantom{-}0.02055661 $ & $\phantom{-}0.02056047$ & $\phantom{-}0.02056176$ & $\phantom{-}0.02056177$ \\[0.05cm]
\hline
 & & & & & \\[-0.3cm]
$\beta^2=\frac{2}{5},\, S^z={\tt w}=1,\, \pi{\tt k}=0.1$ 
& $\phantom{-}2.13319776$ & $\phantom{-}2.11776634$ & $\phantom{-}2.11394936$ & $\phantom{-}2.11267703$ & $\phantom{-}2.11267497$ \\[0.05cm]
\hline
\end{tabular}
}
\vskip 0.15cm
\caption{\small \!\!\! The table gives the quantity $\frac{{\cal K}_p(\bm{w}^{(1)})}{{\cal K}_p(\bm{w}^{(0)})}\times\big(\frac{L}{C}\big)^4$ for a range of $L$ 
and for various values of the parameters. Recall that $2p=\beta\,S^z+\beta^{-1}\,({\tt k}+{\tt w})$. In the second last column,
the numerical data for finite $L$ is interpolated to $L=\infty$. The last column lists $V_p(\bm{w}^{(1)})$, i.e.,
the Virasoro norm \eqref{isiiisaisaias} of the state $L_{-1}\ |v_p\rangle$.
\label{tab1}}
\end{table}
\begin{table}
\centering
\scalebox{0.92}{
\begin{tabular}{|c|c|c|c|c|c|c|}
\hline
 & & & & & & \\[-0.25cm]
 Parameters & State &  $L=50$ & $L=100 $ & $L=200$ & $L=\infty$ & $V_p\big(\bm{w}^{(2,\pm)}\big)$ \\
\hline
 & & & & & & \\[-0.25cm]
\multirow{3}{3cm}{\vskip-0.9cm $S^z=0,\,\beta^2=\frac{20}{43}\vphantom{\Big(}$ ${\tt w}=1,\ \pi{\tt k}=0.1$} & ``$+$'' & $\phantom{-}5.92159778$ & $\phantom{-}5.78605329$ &
$\phantom{-}5.75295649$ & $\phantom{-}5.74192422$ & $\phantom{-}5.74198920$ \\
\cline{2-7}
&  & & & & & \\[-0.25cm]
&  ``$-$'' & $-4.26163068$ & $-4.24034313 $ & $-4.23508161$ & $-4.23332777$ & $-4.23333171$ \\
\hline
&  & & & & & \\[-0.25cm]
\multirow{3}{3cm}{\vskip-0.9cm  $S^z=1,\,\beta^2=\frac{10}{23}\vphantom{\Big(}$ ${\tt w}=0,\, \pi{\tt k}=0.45$}  &  ``$+$'' & $\phantom{-}0.03543829$ & $\phantom{-}0.03533029 $ 
& $\phantom{-}0.03530343$ & $\phantom{-}0.03529447$ & $\phantom{-}0.03529441$ \\
\cline{2-7}
&  & & & & & \\[-0.25cm]
&  ``$-$'' & $-2.66353960$ & $-2.68714565$ & $-2.69308258$ & $-2.69506156$ & $-2.69505911$ \\
\hline
\end{tabular}
}
\vskip 0.2cm
\caption{\small \!\!\!\! \!The table lists numerical values of $\frac{{\cal K}_p(\bm{w}^{(2,\pm)})}{{\cal K}_p(\bm{w}^{(0)})}\times\big(\frac{L}{C}\big)^8$,
 which were computed for two sets of parameters and increasing  $L$ from the Bethe roots via
eqs.\,\eqref{aoso1a},\,\eqref{osissisissd}\,\eqref{ososaoosa} and \eqref{aiusssu}.  The numerical data for the 
constant $C$ is the same as that used in fig.\,\ref{fig6}.
Recall that the sets $\bm{w}^{(2,\pm)}$ label the two basis states $\big|\bm{w}^{(2,\pm)}\big\rangle\!\!\big\rangle$ 
in the level subspace ${\cal F}^{(N)}_p$ with $N=2$, 
given explicitly in eq.\,\eqref{sisisisnn}. 
The data in the ``$L=\infty$'' column was obtained by interpolating the results for finite $L$.
The last column gives $V_p\big(\bm{w}^{(2,\pm)}\big)$ from eq.\,\eqref{oiwoioiuds1}.
\label{tab2}}
\end{table}
\bigskip

The following comment is in order here. In the case when $\Delta=-\frac{1}{2}$ , i.e.,
$\beta^2=\frac{2}{3}$, an explicit analytical expression for the norm $\mathfrak{N}$ 
at finite $L$ was conjectured for certain primary Bethe states 
by Razumov and Stroganov in refs.\cite{Razumov:2000ei,Razumov:2001zg}. In our conventions, 
the expressions proposed in those works translate to
\bea\label{Normformula1}
&&\!\!\!\!\!\!\!\!\!\!\mathfrak{N}^{({\rm odd})}= \big(A(\tfrac{L-1}{2})\big)^4\ 
\prod_{j=1}^{\frac{L-1}{2}}\frac{3(3j-1)^2}{4(2j-1)^2}\qquad \qquad\
\quad\ \ {\rm for} \quad L-{\rm odd}\,,\ \  S^z=\tfrac{1}{2}\,,\ \   {\tt k}=0\,,\ \  {\tt w}=0 \nonumber
 \\[-0.0cm]
&&\\
&&\!\!\!\!\!\!\!\!\!\!\mathfrak{N}^{({\rm even})}=\tfrac{3}{4}\,
\big(A(\tfrac{L}{2})\big)^4\ \prod_{j=1}^{\frac{L}{2}}\frac{4(2j-1)^2\,(3j-1)}{3(3j-2)^3} \ \,
\qquad {\rm for} \quad L-{\rm even}\,,\, \ S^z=0\,,\ \  {\tt k}=\tfrac{1}{3}\,,\  \ {\tt w}=0\,,\nonumber
\eea
where
\be
A(M)=\prod_{j=0}^{M-1}\,\frac{(3j+1)!}{(M+j)!}\nonumber
\ee
is the number of $M\times M$ alternating sign matrices. 
It is straightforward to show that 
the large-$L$ behavior of
\eqref{Normformula1} is consistent with our results
\eqref{aososaik},\,\eqref{ossiss},\,\eqref{isussuus}, provided that
\be
C_0^2\,C^{\frac{1}{18}}=\frac{3^{\frac{11}{36}}\,\re^{\frac{1}{6}}\,\Gamma^2(\frac{1}{3})}{2^{\frac{3}{4}}\,\pi\,A_{{\rm G}}^2}\ \ \ \ \ \ \ \ \ \ \ \ \qquad \Big(\beta=\sqrt{\tfrac{2}{3}}\,\Big)\ .
\ee

\section{Conclusion}
This work is dedicated to 
the description of the scaling behavior of the norms of 
the low energy states of the $XXZ$ spin  chain in the critical regime,
where the anisotropy $-1\leq \Delta<1$.
The key result is the formula \eqref{aososaik} supplemented by the explicit expressions for the constants
${\cal A}_2$\,\eqref{aisiasjjs},
${\cal A}_{1}$\,\eqref{asdasd1122}, the
state dependent exponent $\eta$ \eqref{ossiss} and  the   scaling amplitude ${\mathfrak N}_\infty$ \eqref{isussuus}.
The result  was obtained by  a combination of  analytical techniques based on the ODE/IQFT correspondence
and the numerical analysis of the norms. Currently,  all the above mentioned formulae have a conjectural status.
We believe that their rigorous proof would give a better understanding
of the scaling limit of   integrable lattice models  and, perhaps,  general aspects  of
the RG  flow for  systems in $1+1$ dimensions.
\bigskip

Another result which deserves to be mentioned is 
 the Hermitian conjugation  for  the finite chain given by eqs.\,\eqref{iiisisaas} and \eqref{ssssyys},
which induces the canonical
 conjugation \eqref{sisisiss} of the conformal algebra
in the scaling limit. Clearly, such a modification of the conventional matrix Hermitian structure 
can  be interpreted  as the lattice counterpart of the  Dotsenko-Fateev procedure of introducing the
``charge at infinity'' in the Gaussian model \cite{Dotsenko:1984nm}.
This is of prime importance for the RSOS restrictions
of the 6-vertex model \cite{Andrews:1984af}.
One can note that,  typically, in lattice integrable systems
the Bethe states are not orthogonal w.r.t.  the naive inner product as  the standard matrix  conjugation
does not have any meaningful intrinsic description in the  algebra of 
commuting $Q$- and $T$- operators. An interesting example of such a phenomenon  is provided by the
alternating spin chain associated with the inhomogeneous 6-vertex model \cite{Jacobsen:2005xz}.
In this case  the Hermitian conjugation, which  is consistent with the integrable structure of the model,  becomes
the canonical one for the $W_\infty$-algebra underlying the scaling behavior
of the lattice system. A detailed study of the interplay of  these Hermitian and integrable structures
is of prime importance for understanding the scaling limit of  the alternating spin chain \cite{Ikhlef:2011ay}. Some
results in this direction were reported in the recent work \cite{Bazhanov:2019xvy}.

\section*{Acknowledgments}

The authors thank V. Bazhanov, 
  G. Korchemsky, I. Kostov,
V.~Mangazeev,
 H. Saleur, F. Smirnov and  A. Zamolodchikov for stimulating discussions and important comments.

\medskip
\noindent
The final stage of this work was done during the second author's visit to the IPhT Centre CEA de Saclay. 
SL is grateful to the IPhT for its support and hospitality.

%\newpage
%\section{Appendix}

\end{document}